\documentclass[
    version=last,
    footinclude=false,
    headinclude=false,
    onecolumn,
    oneside,
    paper=A4,
    12pt,
    DIV=11,
    final,
]{scrartcl}

\usepackage[utf8]{inputenc}
\usepackage[main=english]{babel}
\usepackage{libertinus}
\usepackage{csquotes}
\usepackage{enumitem}
\usepackage{microtype}

\usepackage{mathtools}
\mathtoolsset{mathic}
\allowdisplaybreaks[2] 
\usepackage{amsthm}
\usepackage{thmtools}
\usepackage{derivative}
\usepackage[libertinus,slantedGreek,varbb]{newtxmath}
\let\emptyset\varnothing
\usepackage[scr=boondoxo,cal=boondoxo,frak=euler]{mathalpha}

\newcommand\numberthis[1]{\addtocounter{equation}{1}\tag{\theequation}\label{#1}}

\newcommand{\statement}[1]{\paragraph{#1}\pdfbookmark[1]{#1}{#1}} 

\let\epsi\varepsilon

\DeclarePairedDelimiter{\intervaloo}{\lparen}{\rparen}
\DeclarePairedDelimiter{\intervaloc}{\lparen}{\rbrack}

\DeclarePairedDelimiter{\intervalcc}{\lbrack}{\rbrack}

\DeclarePairedDelimiter{\abs}{\lvert}{\rvert}
\DeclarePairedDelimiter{\paren}{\lparen}{\rparen}
\DeclarePairedDelimiter{\fparen}{\lparen}{\rparen}
\DeclarePairedDelimiter{\norm}{\lVert}{\rVert}

    \newcommand{\VERT}[1]{#1|\mkern-1.5mu#1|\mkern-1.5mu#1|}

    \ExplSyntaxOn
    \makeatletter

    \MT_delim_default_inner_wrappers:n{\tnorm}

    \NewDocumentCommand{\tnorm}{ s o m }{
        \IfBooleanTF{#1}{
        \MT_delim_tnorm_star_wrapper:nnn%
            {\VERT{\bgroup\left}}{#3}{\VERT{\aftergroup\egroup\right}}
        }{
            \IfValueTF{#2}{
                \@nameuse{MT_delim_tnorm_nostarscaled_wrapper:nnn}%
                    {\VERT{\@nameuse {\MH_cs_to_str:N #2 l}}}
                    {#3}
                    {\VERT{\@nameuse {\MH_cs_to_str:N #2 r}}}
            }{
                \MT_delim_tnorm_nostarnonscaled_wrapper:nnn%
                    {\VERT{}}
                    {#3}
                    {\VERT{}}
            }
        }
    }

    \makeatother
    \ExplSyntaxOff

\DeclarePairedDelimiter{\commutator}{\lbrack}{\rbrack}
\DeclarePairedDelimiter{\anticommutator}{\{}{\}}
\DeclarePairedDelimiter{\List}{\{}{\}}
\DeclarePairedDelimiter{\ceil}{\lceil}{\rceil}

\DeclarePairedDelimiter{\bra}{\langle}{\rvert}
\DeclarePairedDelimiter{\ket}{\lvert}{\rangle}
\DeclareDocumentCommand{\ketbra}{o m m}{
    \IfValueTF{#1}
        {\ket[#1]{#2}\bra[#1]{#3}}
        {\ket{#2}\bra{#3}}%
}
\DeclareDocumentCommand{\proj}{o m}{
    \IfValueTF{#1}
        {\ket[#1]{#2}\bra[#1]{#2}}
        {\ket{#2}\bra{#2}}%
}

\DeclarePairedDelimiterXPP{\dist}[1]{\operatorname{dist}}{\lparen}{\rparen}{}{#1}
\DeclarePairedDelimiterXPP{\diam}[1]{\operatorname{diam}}{\lparen}{\rparen}{}{#1}
\DeclarePairedDelimiterXPP{\Ld}[1]{d}{\lparen}{\rparen}{}{\ifblank{#1}{\interpunct,\interpunct}{#1}}
\DeclarePairedDelimiterXPP{\Ldist}[1]{d}{\lparen}{\rparen}{}{#1}
\DeclarePairedDelimiterXPP{\Ldiam}[1]{\SYMdiam}{\lparen}{\rparen}{}{#1}

\makeatletter
\let\bBigg@@\bBigg@
\renewcommand{\bBigg@}[2]{{%
  \mathchoice
    {\bBigg@@{#1}{#2}}%
    {\bBigg@@{#1}{#2}}%
    {\big@size=.5\big@size\bBigg@@{#1}{#2}}%
    {\big@size=.3\big@size\bBigg@@{#1}{#2}}}}%
\makeatother

\DeclareDocumentCommand{\trace}{s o e{_} m}{
    \IfValueTF{#3}
        {\Tr_{#3}}
        {\Tr}%
    \IfBooleanTF{#1}
        {\paren*{#4}}
        {
            \IfValueTF{#2}
                {\paren[#2]{#4}}
                {\paren{#4}}%
        }%
}
\DeclareDocumentCommand{\Trace}{s o e{_} m}{
    \IfValueTF{#3}
        {\Tr_{#3}}
        {\Tr}%
    \IfBooleanTF{#1}
        {\paren*{#4}}
        {
            \IfValueTF{#2}
                {\paren[#2]{#4}}
                {\paren{#4}}%
        }%
}

\providecommand\given{}

\newcommand\SetSymbol[1][]{%
    \nonscript\,#1\vert
    \allowbreak
    \nonscript\,
    \mathopen{}}
\DeclarePairedDelimiterX\Set[1]\{\}{%
    \renewcommand\given{%
        \SetSymbol[\delimsize]}
    \nonscript\,
    #1
    \nonscript\,
}

\ExplSyntaxOn
\DeclareDocumentCommand{\cexp}{s o m m}{%
    \cexpsym\c_math_subscript_token{#3}
    \IfBlankF{#4}
    {
        \exp_last_unbraced:Ne \paren {\IfBooleanT{#1}{*}\IfValueT{#2}{[\exp_not:N #2]}} {#4}
    }
}
\ExplSyntaxOff

\ExplSyntaxOn
\makeatletter
\cs_generate_variant:Nn \tl_if_eq:nnT {onT}
\tl_new:N \__scale_new_delimsize_tl
\cs_new:Nn \__scale_make_bigger_tl:N {
    \tl_if_eq:onT {#1} {\@MHempty} { \tl_set:Nn \__scale_new_delimsize_tl { big } }
    \tl_if_eq:onT {#1} {}          { \tl_set:Nn \__scale_new_delimsize_tl { big } }
    \tl_if_eq:onT {#1} {\big}      { \tl_set:Nn \__scale_new_delimsize_tl { Big } }
    \tl_if_eq:onT {#1} {\Big}      { \tl_set:Nn \__scale_new_delimsize_tl { bigg } }
    \tl_if_eq:onT {#1} {\bigg}     { \tl_set:Nn \__scale_new_delimsize_tl { Bigg } }
    \tl_if_eq:onT {#1} {\middle}   { \tl_set:Nn \__scale_new_delimsize_tl { middle } }
}
\cs_new:Nn \scale_make_bigger_m:N {
    \__scale_make_bigger_tl:N #1
    \use:c { \tl_use:N \__scale_new_delimsize_tl }
}
\cs_new:Nn \scale_make_bigger_l:N {
    \__scale_make_bigger_tl:N #1
    \tl_if_eq:NnTF \__scale_new_delimsize_tl {middle} {
        \tl_set:Nn \__scale_new_delimsize_tl {left}
    }{
        \tl_put_right:Nn \__scale_new_delimsize_tl {l}
    }
    \use:c { \tl_use:N \__scale_new_delimsize_tl }
}
\cs_new:Nn \scale_make_bigger_r:N {
    \__scale_make_bigger_tl:N #1
    \tl_if_eq:NnTF \__scale_new_delimsize_tl {middle} {
        \tl_set:Nn \__scale_new_delimsize_tl {right}
    }{
        \tl_put_right:Nn \__scale_new_delimsize_tl {r}
    }
    \use:c { \tl_use:N \__scale_new_delimsize_tl }
}
\DeclarePairedDelimiterXPP{\pLd}[1]{\scale_make_bigger_l:N\delimsize\lparen d}{\lparen}{\rparen}{\scale_make_bigger_r:N\delimsize\rparen}{#1}
\DeclarePairedDelimiterXPP{\pLdist}[1]{\scale_make_bigger_l:N\delimsize\lparen d}{\lparen}{\rparen}{\scale_make_bigger_r:N\delimsize\rparen}{#1}
\DeclarePairedDelimiterXPP{\pLdiam}[1]{\scale_make_bigger_l:N\delimsize\lparen \SYMdiam}{\lparen}{\rparen}{\scale_make_bigger_r:N\delimsize\rparen}{#1}

\makeatother
\ExplSyntaxOff

\usepackage[svgnames]{xcolor}

\usepackage[
    hidelinks,
    bookmarksnumbered,
    bookmarksopen,
    bookmarksopenlevel=2,
    bookmarksdepth=3,
    pdfusetitle=true,
    breaklinks=true,
]{hyperref}

\makeatletter
\@ifpackageloaded{hyperref}{
    \usepackage[capitalise,noabbrev]{cleveref}
    \usepackage{orcidlink}
}{
    \usepackage[capitalise,noabbrev,poorman]{cleveref}
    \newcommand{\texorpdfstring}[2]{#1}
    \newcommand{\href}[2]{#2}
    \newcommand{\hypersetup}[1]{}
    \newcommand{\orcidlink}[1]{ORCiD}
    \newcommand{\pdfbookmark}[1]{}
}
\makeatother

\crefdefaultlabelformat{#2#1#3}
\crefname{equation}{}{}
\crefname{assumption}{Assumption}{Assumptions}
\crefname{conjecture}{Conjecture}{Conjectures}
\renewcommand{\namecref}{\lcnamecref}

\usepackage{caption}
\usepackage{subcaption}
\usepackage{floatrow}
\makeatletter
\newcommand*{\diff}{\@ifnextchar^{\DIfF}{\DIfF^{}}}
\def\DIfF^#1{%
    \mathop{\mathrm{\mathstrut d}}\nolimits^{#1}\gobblespace}
\def\gobblespace{\futurelet\diffarg\opspace}
\def\opspace{%
    \let\DiffSpace\!%
    \ifx\diffarg(%
        \let\DiffSpace\relax
    \else
        \ifx\diffarg[%
            \let\DiffSpace\relax
        \else
            \ifx\diffarg\{%
                \let\DiffSpace\relax
            \fi
        \fi
    \fi
    \DiffSpace
} 
\makeatother

\newcommand{\sumstack}[2][]{\ifstrempty{#1}{\sum_{\substack{#2}}}{\smashoperator[#1]{\sum_{\substack{#2}}}}}

\newcommand{\e}{{\mathrm{e}}}
\newcommand{\I}{\mathrm{i}}
\newcommand{\N}{\mathbb{N}}
\newcommand{\Z}{\mathbb{Z}}
\newcommand{\R}{\mathbb{R}}
\newcommand{\C}{\mathbb{C}}

\newcommand{\HS}{{\mathcal{H}}}
\newcommand{\fock}{\symcal{F}}

\newcommand{\numberOp}{{\symcal{N}}}
\newcommand{\even}{^+}

\newcommand{\alg}{{\symcal{A}}}
\newcommand{\algN}{\alg^{\numberOp}}
\newcommand{\algEven}{\alg\even}

\newcommand{\unit}{\mathbf{1}}

\makeatletter
\NewDocumentCommand{\uts}{ O{t,s} D<>{} m O{}}{%
    \tau^{#2}_{#1\ifstrempty{#4}{}{,#4}}\ifstrempty{#3}{}{\fparen{#3}}}
\makeatother
\DeclareDocumentCommand{\op}{s m}{
    \IfBooleanTF{#1}{a^*_{#2}}{a^{}_{#2}}
}

\DeclareMathOperator{\id}{id}

\newcommand{\boundedOps}{{\symcal{B}}}

\newcommand{\interactions}[1]{\symcal{S}_{#1}}

\newcommand{\Ldim}{D}

\newcommand{\Lat}[1][]{\Lambda_{#1}}

\newcommand{\Wf}{\symcal{W}}
\newcommand{\Wfgd}{\symcal{W}\kern-4pt_{g,\delta}}
\newcommand{\ftWf}{\widehat{\Wf}}

\newcommand{\qliL}{\symcal{I}}

\newcommand*{\velocity}{\mathscr{v}}
\newcommand{\Vconst}{\symcal{C}_\mathup{V}}
\newcommand{\Aconst}{\symcal{C}_\mathup{A}}
\newcommand{\LRconst}{\symcal{C}_\mathup{LR}}
\newcommand{\graphs}[1][\Ldim,\Aconst]{\mathcal{G}(#1)}
\newcommand{\sigmaconst}{C_{\sigma}}
\newcommand{\latticesumconst}{\symcal{C}^{\textup{Lem.\,\ref{lem:bound-lattice-sum-by-integral}}}}
\newcommand{\INTconst}[1][\mu,\nu]{\symcal{C}_{#1}^{\textup{Lem.\,\ref{lem:bound-integration-exp-times-polynomial}}}}
\newcommand{\LRint}{\symcal{I}}
\newcommand{\spin}{{\symcal{s}}}
\newcommand{\us}{^{\mathrm{spin}}}
\newcommand{\sltoperator}{K}
\newcommand{\qlilgenerator}{G}
\newcommand{\bSets}[2][]{S_{#1}(#2)}
\newcommand{\LRBoundarySet}[2]{#1\subset\bSets{#2}}
\newcommand{\alphatb}{\alpha_{\mathup{tb}}}

\newcommand{\cexpsym}{\mathbb{E}}
\newcommand{\SYMdiam}{\operatorname{diam}}
\newcommand{\GammaFunc}{\upGamma}

\DeclareMathOperator{\Tr}{Tr}

\newcommand{\suchthat}{\mathpunct{\ordinarycolon}}

\newcommand{\quadtext}[1]{\quad\text{#1}\quad}
\newcommand{\qquadtext}[1]{\quad\quadtext{#1}\quad}
\newcommand{\hspacearound}[2]{\hspace{#1}#2\hspace{#1}}
\newcommand{\interpunct}{\mathop{\cdot}}

\newcommand{\Alignindent}{\hspace*{2em}&\hspace*{-2em}}
\providecommand{\symcal}[1]{\mathcal{#1}}

\providecommand{\mathup}[1]{\mathrm{#1}}

\let\oldsetminus\setminus
\newbox\mybox
\newcommand\cutsetminus[1]{%
    \setbox\mybox\hbox{\(#1\oldsetminus\)}%
    \ht\mybox=0pt%
    \dp\mybox=0pt%
    \usebox\mybox%
}
\renewcommand\setminus{%
    \mathbin{%
        \mathchoice%
            {\displaystyle\oldsetminus}
            {\textstyle\oldsetminus}
            {\cutsetminus{\scriptstyle}}
            {\cutsetminus{\scriptscriptstyle}}
    }%
}

\makeatletter
\ExplSyntaxOn

\newcounter{theoremenv}
\newcounter{theoremenvglobal}
\newlist{thmlist}{enumerate}{1}
\setlist[thmlist]{
    label=\textup{(\alph{thmlisti})},
    ref={(\alph{thmlisti})},
    nosep,
}
\renewcommand{\p@thmlisti}{\perh@ps{\protect\ref{auto-label:\arabic{theoremenvglobal}}}}
\DeclareRobustCommand{\perh@ps}[1]{#1}
\newcommand{\itemref}[1]{%
    \begingroup 
    \let\perh@ps\@gobble\ref{#1}%
    \endgroup
}

\addtotheorempostheadhook{
    \crefalias{thmlisti}{\thmt@envname}
    \setcounter{theoremenv}{\value{\thmt@envname}}
    \stepcounter{theoremenvglobal}
    \exp_args:NNe \renewcommand \thetheoremenv {\use:c{the\thmt@envname}}
    \exp_args:Ne \label {auto-label:\arabic{theoremenvglobal}}
}
\makeatother
\ExplSyntaxOff

\theoremstyle{plain}

\declaretheorem[
    name=Theorem,
]{theorem}

\declaretheorem[
    name=Lemma,
    sibling=theorem,
]{lemma}

\declaretheorem[
    name=Proposition,
    sibling=theorem,
]{proposition}

\declaretheorem[
    name=Corollary,
    sibling=theorem,
]{corollary}

\theoremstyle{definition}

\theoremstyle{remark}
\declaretheorem[
    name=Remark,
    sibling=theorem,
    qed=\(\diamond\),
]{remark}

\declaretheorem[
    name=Assumption,
    style=definition,
    sibling=theorem,
]{assumption}

\usepackage{tikz}
\newlength{\ul}
\newlength{\lw}
\definecolorset{HTML}{col_}{}{%
    1,0072B2;%
    2,D55E00;%
    3,E69F00
}

\usetikzlibrary{
    arrows.meta,
    calc,
    fit,
    intersections,
    positioning,
    shapes,
    through,
}

\tikzset{%
    myLine/.style = {
        line width = #1\lw,
        draw = black,
    },
    myLine/.default = 1,
    x = \ul,
    y = \ul,
    text = black,
    >={Straight Barb},
    shorten/.style = {
        shorten < = #1,
        shorten > = #1
    },
    shorten/.default = 2\lw,
    set/.style = {
        shape = ellipse,
        myLine,
        minimum height = 0.4\ul,
        minimum width = 1\ul,
        fill = #1!20,
        draw = #1,
        text = #1,
        inner sep = .1\lw,
        outer sep = 0pt,
        x = \ul,
        y = \ul,
    },
    set/.default = gray,
    bounding note/.style = {
        use as bounding box,
        shape = rectangle,
        inner sep = 0pt,
        outer sep = 0pt,
    },
}

\ExplSyntaxOn
\makeatletter
\def\@altpic{0}
\newcommand\altpic[3]{
\ifx\@altpic\undefined
    \PackageError{custom}{step is undefined}{} undefined
\else
    \if\@altpic1
        #1
    \else
        \if\@altpic2
        #2
        \else
            \if\@altpic3
            #3
            \else
            \PackageError{custom}{step~is~not~1~or~2~or~3}{}
            other
            \fi
        \fi
    \fi
\fi
}
\newcommand\setaltpic[1]{\def\@altpic{#1}}

\makeatother
\ExplSyntaxOff

\usepackage[
    bibstyle=phys,
    citestyle=numeric,
    biblabel=brackets,
    mincitenames=1,
    maxcitenames=3,
    minalphanames=3,
    maxalphanames=4,
    minbibnames=3,
    maxbibnames=10,
    arxiv=abs,
    eprint=true,
    doi=false,
    url=false,
    sorting=nty,
    useprefix=true,
    alldates=year,
    urldate=long,
]{biblatex}
\renewbibmacro*{note+pages}{%
    \printfield{note}%
    \setunit{\bibpagespunct}%
    \printfield{pages}%
    \newunit%
}
\makeatletter
\DeclareFieldFormat{eprint:arxiv}{%
    \ifhyperref
    {\href{https://arxiv.org/\abx@arxivpath/#1}{%
            arXiv\addcolon\linebreak[2]#1}}
    {arXiv\addcolon
        \nolinkurl{#1}%
    }
}
\makeatother

\renewbibmacro*{maintitle+title}{%
    \iffieldsequal{maintitle}{title}{%
        \clearfield{maintitle}%
        \clearfield{mainsubtitle}%
        \clearfield{maintitleaddon}%
    }{%
        \iffieldundef{maintitle}{}{%
            \printtext[doi/url-link]{%
                \usebibmacro{maintitle}%
            }%
            \newunit\newblock
            \iffieldundef{volume}{}{%
                \printfield{volume}%
                \printfield{part}%
                \setunit{\addcolon\space}%
            }%
        }%
    }%
    \printtext[doi/url-link]{%
        \usebibmacro{title}%
    }%
    \newunit%
}

\DeclareBibliographyDriver{article}{%
    \usebibmacro{bibindex}%
    \usebibmacro{begentry}%
    \usebibmacro{author/translator+others}%
    \setunit{\labelnamepunct}\newblock
    \usebibmacro{title}%
    \newunit
    \printlist{language}%
    \newunit\newblock
    \usebibmacro{byauthor}%
    \newunit\newblock
    \usebibmacro{bytranslator+others}%
    \newunit\newblock
    \printfield{version}%
    \newunit\newblock
    \printtext[doi/url-link]{%
        \usebibmacro{journal+issuetitle}%
        \newunit
        \usebibmacro{byeditor+others}%
        \newunit
        \usebibmacro{note+pages}%
        \newunit\newblock
        \iftoggle{bbx:isbn}
        {\printfield{issn}}
        {}%
    }%
    \newunit\newblock
    \printfield{pubstate}%
    \setunit{\addspace}%
    \printfield{year}%
    \newunit\newblock
    \usebibmacro{doi+eprint+url}%
    \newunit\newblock
    \printfield{addendum}%
    \setunit{\bibpagerefpunct}\newblock
    \usebibmacro{pageref}%
    \newunit\newblock
    \iftoggle{bbx:related}
        {\usebibmacro{related:init}%
        \usebibmacro{related}}
        {}%
    \usebibmacro{finentry}%
}

\usepackage{xurl} 

\KOMAoptions{toc=bib}
\setkomafont{sectionentry}{\normalfont\normalsize\sffamily}
\setkomafont{author}{\sffamily}
\setkomafont{date}{\sffamily}
\DeclareTOCStyleEntry[level:=section,indent:=section,beforeskip=0pt,linefill=\TOCLineLeaderFill,numwidth:=section,pagenumberwidth:=section]{tocline}{section}
\date{18. December 2025}

\title{Lieb-Robinson bounds, automorphic equivalence and LPPL for long-range interacting fermions}

\author{
    Stefan Teufel%
    \texorpdfstring{%
        \,\orcidlink{0000-0003-3296-4261}
        \footnote{
            \parbox[t]{.75\textwidth}{
                \foreignlanguage{ngerman}{Fachbereich Mathematik, Universität~Tübingen,\\
                Auf~der~Morgenstelle~10, 72076~Tübingen,} Germany
                \\Email:
                \href{mailto:stefan.teufel@uni-tuebingen.de}{stefan.teufel@uni-tuebingen.de},
                \href{mailto:tom.wessel@uni-tuebingen.de}{tom.wessel@uni-tuebingen.de}
            }
        }
    }{}%
    \and
    Tom Wessel%
    \texorpdfstring{%
        \,\orcidlink{0000-0001-7593-0913}
        \footnotemark[1]
    }{}%
}

\begin{document}

\maketitle

\begin{abstract}
    We prove a Lieb-Robinson bound for lattice fermion models with polynomially decaying interactions, which can be used to show the locality of the quasi-local inverse Liouvillian.
    This allows us to prove automorphic equivalence and the local perturbations perturb locally (LPPL) principle for these systems.
    The proof of the Lieb-Robinson bound is based on the work of \textcite{EMN2020}, and our results also apply to spin systems.
    We explain why some newer Lieb-Robinson bounds for long-range spin systems cannot be used to prove the locality of the quasi-local inverse Liouvillian, and in some cases may not even hold for fermionic systems.
\end{abstract}

\tableofcontents
\newpage

\section{Introduction}
\label{sec:Introduction}

Locality is an important feature of many-body quantum systems.
And a key tool to characterize locality in quantum lattice systems are Lieb-Robinson bounds~\cite{LR1972}, which underlie the existence of the thermodynamic limit of the Heisenberg dynamics~\cite{LR1972,NOS2006}, decay of correlations for gapped ground states~\cite{HK2006,NS2006,WH2022}, the characterization of topological phases~\cite{HW2005,BMNS2012}, the adiabatic theorem~\cite{BRF2018,MT2019} and generalized response theory~\cite{Teufel2020,HT2020bulk}.

Lieb-Robinson bounds measure locality in lattice systems, where the Hamiltonian is given as a sum of local terms
\begin{equation*}
    H
    =
    \sum_{Z\subset \Lambda}
    \Phi(Z)
    ,
\end{equation*}
where each \(\Phi(Z)\) is only acting on the sites in \(Z\).
Here, \(\Lambda\) is a finite lattice and the Hilbert space could either describe a spin system, which means at each site there are some spin degrees of freedom, or a fermionic lattice system, describing fermions moving on the lattice.
Denoting with \(\tau_{t}\) the time evolution generated by \(H\), a Lieb-Robinson bound is an upper bound for
\begin{equation*}
    \norm{
        \commutator{
            \tau_t(A), B
        }
    }
    ,
\end{equation*}
where the observables \(A\) and \(B\) are supported on disjoint sets \(X\) and \(Y\subset \Lambda\), respectively, and, for fermionic systems, at least one of them is even.
As the commutator would exactly vanish if \(\tau_t(A)\) was supported in \(\Lambda \setminus Y\), Lieb-Robinson bounds provide a measure of locality of \(\tau_t(A)\).

For \emph{short-range} interactions, i.e.\ if \(\norm{\Phi(Z)}\) decays exponentially in \(\Ldiam{Z}\), the Lieb-Robinson bounds have the form~\cite{HK2006,NS2006,NS2009,NSY2019}
\begin{equation}
    \label{eq:LRshort}
    \norm{
        \commutator{
            \tau_t(A), B
        }
    }
    \leq
    C
    \, \min\List{\abs{X},\abs{Y}}
    \, \norm{A} \, \norm{B}
    \, \e^{b(\velocity t-r)}
    .
\end{equation}
This implies, in particular, that the time-evolved operator is supported inside a ball of radius \(r\leq\velocity t\), up to exponentially decaying terms outside.
Accordingly, this region is often called the \emph{light cone}, and more specifically the Lieb-Robinson bound for short-range interactions~\eqref{eq:LRshort} exhibits a \emph{linear light cone}, as the bound is small up to times \(\velocity t \sim r\).

The main purpose of this paper is to prove locality for the quasi-local inverse Liouvillian and automorphic equivalence of gapped ground states in the same phase, which we explain later, for \emph{long-range} interactions, where \(\norm{\Phi(Z)}\) decays polynomially in \(\Ldiam{Z}\), in fermionic lattice systems.
The standard proofs of Lieb-Robinson bounds also give bounds for long-range interactions~\cite{HK2006}, but they only yield a logarithmic light cone and are not strong enough to prove automorphic equivalence~\cite{NSY2019}.
Recently, there was a focus on proving Lieb-Robinson bounds with linear light cones for very slow polynomial decay~\cite{FGC2015,KS2020,TCE2020,TGB2021}.
However, these works typically focus on large times and two-body interactions, making them not suitable for our application.
We need to apply the Lieb-Robinson bound to the generator of the spectral flow, which contains general many-body interactions, and we need the bounds also for small times.
Moreover, as we explain in \cref{sec:comments-on-conditional-expectation}, some of the proofs~\cite{TCE2020,TGB2021} only work for spin systems but not for fermions.

Hence, we focus on two results~\cite{MKN2016,EMN2020} for long-range interactions, with comparably short proofs, which can easily be generalized to fermionic systems.
These results have a root-like light cone, but that is no problem for our application.
For our long-term goal of proving the adiabatic theorem and generalized response theory, these bounds have a problematic scaling in the size of \(\abs{X}\), which unlike the other cases appears with a possibly high power.
To mitigate this, we first improve these results to the common linear prefactor \(\abs{X}\), see \cref{thm:LR-bound-after-iteration}.

Based on the improved Lieb-Robinson bounds, we prove locality of the quasi-local inverse Liouvillian and automorphic equivalence in these systems, following the corresponding proof strategies for short-range interactions~\cite{HW2005,BMN2011,NSY2017,BRF2018}.
For simplicity, we restrict to two-body interactions on the square lattice to explain some of these results here, while the main text covers arbitrary, finite, surface-regular graphs and many-body interactions.
For the moment, let \(\Lambda\subset \Z^D\) be a finite lattice and \(s\mapsto H(s) = \sum_{x,y\in \Lambda} \Phi(\List{x,y},s)\) be a smooth family of gapped Hamiltonians comprised of two-body interactions with decay \(\norm{\Phi(\List{x,y},s)} \leq C \, (1+\Ldist{x,y})^{\alphatb}\) for some%
\footnote{%
    As we restrict to two-body interactions here, we follow the convention that \(\alphatb\) characterises the decay of each two-body term.
    The power \(\alpha\) used in the main text, which appears in the definition of the interaction norm~\eqref{eq:def-interaction-norm}, has a different meaning.
    See the discussion after \cref{cor:LPPL}.
}
\(\alphatb>3D+1\), and similar decay for \(\dot \Phi\).
Then, an automorphism that connects the instantaneous ground states exists.
It is generated by an interaction with a sufficiently high-power polynomial decay and still satisfies Lieb-Robinson bounds.
Our result immediately implies automorphic equivalence for gapped ground states of interactions with super-polynomial decay, with the automorphisms being generated by interactions with super-polynomial decay as well.
These Lieb-Robinson bounds are also used to prove automorphic equivalence in~\cite{BTW2025automorphic} and an adiabatic theorem in~\cite{BTW2025generalized} for infinite volume systems with super-polynomially decaying interactions and a gap in the bulk.
Moreover, they might be useful for proving stability of the gap for super-polynomially decaying interactions, as implicitly suggested by the comments in~\cite[Appendix~E]{NSY2019}.

Automorphic equivalence can also be shown using the known bounds from~\cite{EMN2020}, albeit with slightly different assumptions.
But repeated applications of the inverse Liouvillian, as needed in the proof of super-adiabatic theorems~\cite{BRF2018,MT2019,Teufel2020}, yield a much worse decay with the Lieb-Robinson bound of~\cite{EMN2020}.
See the discussion after \cref{thm:automorphic-equivalence}.
Unfortunately, even with the current Lieb-Robinson bounds and their interplay with various other steps in the proofs, one still requires very high power-law decay to obtain super-adiabatic theorems.
Hence, we leave this discussion until further improvements of the Lieb-Robinson bounds%
\footnote{
    We later argue that an improved light cone with same spatial decay as achieved in~\cite{TGB2021} does not improve the results on automorphic equivalence in \cref{thm:automorphic-equivalence}.
    However, it can still improve other necessary steps in the proofs of generalized super-adiabatic theorems.
}
and the involved steps are developed.

Using the same ideas underlying automorphic equivalence, we also prove the local perturbations perturb locally principle for these systems, see \cref{cor:LPPL}.
Again as a simplified special case, consider \(H(s) = \sum_{x,y\in \Lambda} \Phi(\List{x,y}) + s \, W\) for some two-body interaction \(\Phi\) with decay exponent \(\alphatb > 2D\) and \(W\) supported on \(X\subset \Lambda\).
Moreover, assume that the ground state of the Hamiltonian \(H(s)\) is gapped, with the gap bounded below uniformly in \(s\) and denote with \(P(s)\) the ground state projection.
Then, for every \(\beta<\alphatb-D\) and observable \(A\) supported on \(Y\subset \Lambda\),
\begin{equation*}
    \abs[\big]{ \trace[\big]{P(s)\,A} -\trace[\big]{P(0)\,A} }
    \lesssim
    \norm{A} \, \abs{Y} \, \norm{\dot H(t)} \, \paren[\big]{\Ldist{X,Y}+1}^{-\beta}
    .
\end{equation*}
While our result applies also to general many-body interactions, we note that a different strategy was used in~\cite{WH2022} to obtain a similar LPPL statement with better decay, albeit only for two-body interactions.
See the discussion after \cref{cor:LPPL} for details.

The main text is organized as follows.
In \cref{sec:Mathematical-framework} we discuss the lattices, operator algebras and interaction norms we use.
Afterwards we state the Lieb-Robinson bounds in \cref{sec:Improved-LR-bounds} including the proof of our main improvement.
For completeness and convenience of the reader, all intermediate steps which are similar to previous results are proven in the appendix.
Using the Lieb-Robinson bounds, we prove automorphic equivalence in \cref{sec:Automorphic-equivalence} and discuss LPPL in \cref{sec:LPPL}.
Eventually, we discuss why a trick commonly used for spin systems, which would simplify our proof and underlies the sharpest Lieb-Robinson bounds~\cite{TCE2020,TGB2021}, seems to not work for fermionic systems.

\section{Mathematical framework}
\label{sec:Mathematical-framework}

This section introduces the mathematical framework that we use to describe long-range interacting fermions on a lattice.

\subsection{Lattice and operator algebras}
\label{sec:Lattice-and-operator-algebras}

As lattices, we consider finite graphs \((\Lat,E)\) with graph distance \(\Ld{}\) such that there exist constants \(\Ldim\in{\N}_+\) and \(\Aconst \geq 1\) such that
\begin{equation}
    \abs[\big]{S_y(R)}
    := \abs[\big]{\Set{x\in\Lat \given \Ld{y,x}=R}}
    \leq \Aconst \, R^{\Ldim-1}
    \quad\text{for all \(y\in\Lat\) and \(R\geq1\)}.
    \label{eq:condition-lattice-volume-sphere}
\end{equation}
These graphs are called \emph{surface-regular}, and the set of all \(D\)-dimensional surface-regular graphs with growth constant~\(\Aconst\) is denoted~\(\graphs\).
As the graphs are assumed to be finite, the existence of these constants is trivial.
The crucial point is that all bounds we derive will be uniform for all graphs in the class \(\graphs\), i.e.\ they only depend on the dimension \(D\) and the growth constant \(\Aconst\), but not the specific~\(\Lambda\).
In particular, they are independent of~\(\abs{\Lambda}\).
Hence, one can always take an increasing sequence \(\Lat[k]\subset\Lat[k+1]\) and obtain bounds independent of \(k\) whenever \(\Aconst\) and \(\Ldim\) are uniformly bounded in \(k\).

All the following definitions implicitly depend on the specific lattice \(\Lat\in\graphs\) which will be clear from the context later on.
If \(\Lat\) is not explicitly specified, the statements and definitions hold for all \(D\in{\N}_+\), \(\Aconst>0\) and \(\Lat\in\graphs\).

A simple integration shows that every surface-regular graph is also a regular graph in the sense that
\begin{equation}
    \label{eq:condition-lattice-volume-ball}
    \abs[\big]{B_y(R)}
    := \abs[\big]{\Set{x\in\Lat \given \Ld{y,x}\leq R}}
    \leq \Vconst \, (R+1)^\Ldim
    \quad\text{for all \(y\in\Lat\) and \(R\geq0\)}
    ,
\end{equation}
with \(\Vconst\leq\max\List{1,\Aconst/\Ldim}\).
One could use the more general class of regular graphs, but all the examples we have in mind are surface-regular and one obtains stronger results under this restriction, compare~\cite{MKN2016}.

The above conditions in particular include the square lattices
\begin{equation*}
    \Lat := \List{-k,\dotsc,k}^\Ldim \subset {\Z}^\Ldim
    \quad\text{for some \(k\in{\N}\)}
\end{equation*}
with \(\ell^1\)-metric and additionally allow for torus or cylinder geometries.
In these geometries, points on opposite ends of the lattice \(\Lat\), would have distance \(1\).

We use \(\Ld{}\) also to denote the distance \(\Ldist{x,Y} := \inf_{y\in Y} \Ld{x,y}\) between a point \(x\in \Lambda\) and a set \(Y\subset \Lambda\) and \(\Ldist{X,Y} := \inf_{x\in X, y\in Y} \Ld{x,y}\) between two sets \(X\), \(Y\subset \Lambda\).
We denote with
\begin{equation*}
    \Ldiam{X}
    :=
    \sup \,\Set[\big]{\Ld{x,y} \given x, y\in X }
\end{equation*}
the diameter and with \(\abs{X}\) the cardinality of any set \(X\subset\Lat\).
Furthermore, for \(X\subset\Lat\) and \(m\geq0\), denote by \(X_m\) the \emph{fattening}
\begin{equation}
    \label{eq:definition-fattening}
    X_m := \Set[\big]{y\in\Lat{} \given \Ldist{y,X}\leq m}
    .
\end{equation}

\subsection{Operator Algebra for fermions}
\label{sec:operator-algebra}

For a fermion with spin \(\spin\in{\N}\) on the lattice \(\Lat\), the one particle Hilbert space is \({\HS:=\ell^2(\Lat,{\C}^\spin)}\).
For \(N\in{\N_+}\), the \(N\)-particle Hilbert space is the antisymmetric tensor product \({\HS_{N} := \bigwedge_{k=1}^N \HS}\), and the fermionic Fock space is \(\fock:=\bigoplus_{N=0}^{\spin\,\abs{\Lat}} \HS_{N}\), where \(\HS_{0} := {\C}\).
These Hilbert spaces are all finite-dimensional and hence all linear operators on them are bounded.
The C*-algebra \(\alg_\Lambda=\boundedOps(\fock)\) of bounded operators on \(\fock\) is generated by the fermionic annihilation and creation operators \(\op{z,i}\) and \(\op*{z,i}\).
They satisfy the canonical anti-commutation relations (CAR)
\begin{equation*}
    \anticommutator[\big]{\op{x,i},\op{y,j}}=0=\anticommutator[\big]{\op*{x,i},\op*{y,j}}
    \quadtext{and}
    \anticommutator[\big]{\op{x,i},\op*{y,j}}=\delta_{i,j}\,\delta_{x,y}\,\unit
\end{equation*}
for all \(i,j\in\{1,\dotsc,\spin\}\) and \(x,y\in\Lambda\), where \(\anticommutator{A,B}:=A\,B+B\,A\) denotes the anti-commutator.
For each subset \(Z\subset\Lambda\) one defines the subalgebra \(\alg_Z\) as the one generated by annihilation and creation operators supported in \(Z\).
This yields a natural inclusion \(\alg_{Z'}\subset\alg_Z\) for \(Z'\subset Z\subset\Lat\).

By \(\algEven_{Z}\subset\alg_Z\) we denote the subalgebra generated by products of an even number of creation and annihilation operators located in \(Z\).
For disjoint subsets \(X,Y\subset\Lat\) and operators \(A\in\algEven_{X}\), \(B\in\alg_Y\), it then holds that \(\commutator{A,B}=0\) by the CAR\@.
Moreover, if \(A\in\alg_X\), \(B\in\alg_Y\) and \(\commutator{A,B}=0\), then \(A\in\algEven_{X}\) or \(B\in\algEven_{Y}\) must hold true~\cite[Proposition~2.1]{NSY2017}.

The restriction to even operators is natural for most applications, because in standard physical models no single fermionic particles are created or annihilated.
For our purpose, we are actually often interested in operators conserving the particle number.
And indeed, the subset \(\algN_Z\subset\alg_Z\) of elements commuting with the number operator
\begin{equation*}
    \numberOp_Z
    =
    \sum_{z\in Z}
    \sum_{i=1}^\spin
    \op*{z,i} \, \op{z,i}
\end{equation*}
is a subalgebra \(\algN_Z\subset\algEven_{Z}\) of the even algebra.

\subsection{Interactions and operator-families}
\label{sec:Interactions-and-operator-families}

An \emph{interaction} is a function
\begin{equation*}
    \Phi\colon \Set{Z\subset\Lat} \rightarrow \algEven_{\Lat},\quad
    Z\mapsto\Phi(Z)\in\algEven_Z
    \quadtext{where}
    \Phi(Z)=\Phi(Z)^* \text{ for all \(Z\subset\Lat\)}
    .
\end{equation*}
We denote the space of interactions defined on \(\Lambda\) as \(\interactions{\Lambda}\).
Associated to each interaction is a \emph{sum-of-local-terms}~(SLT) operator
\begin{equation*}
    S := \sum_{Z\subset\Lat} \Phi(Z)
    .
\end{equation*}
Note that each interaction term is assumed to be even and self-adjoint.
Hence, also the corresponding SLT operator is even and self-adjoint.

To control the locality of an interaction, we use the \emph{interaction norm}
\begin{equation}
    \label{eq:def-interaction-norm}
    \norm{\Phi}_{\alpha,n}
    :=
    \sup_{z\in\Lat} \sumstack{Z\subset\Lat\suchthat\\z\in Z} \abs{Z}^n \, \frac{\norm{\Phi(Z)}}{F_\alpha\pLdiam{Z}}
    ,
\end{equation}
where \(n\in{\N}\), \(\alpha\geq0\) and \(F_\alpha(r):= (r+1)^{-\alpha}\), and we write \(\norm{\interpunct}_{\alpha} := \norm{\interpunct}_{\alpha,0}\) for short.

For an interval \(I\subset{\R}\), \emph{time-dependent interactions} are functions \(\Phi(\cdot, \cdot)\) where \(\Phi(\cdot, t)\) is an interaction for each \(t\in I\) and \(\Phi(Z,\cdot)\) is norm continuous for all \(Z\subset\Lat\).
We denote the space of time-dependent interactions defined on \(\Lambda\) for an interval \(I\) as \(\interactions{\Lambda}(I)\).
For \(\Phi\in \interactions{\Lambda}(I)\) we write \(\mathord{\norm{\Phi(t)}_{\alpha,n}}:=\norm{\Phi(\cdot,t)}_{\alpha,n}\) and define
\begin{equation}
    \norm{\Phi}_{\alpha,n}
    :=
    \sup_{t\in I} \mathord{\norm{\Phi(t)}_{\alpha,n}}
    .
\end{equation}

Furthermore, for any function \(F\colon \N \to \R_{\geq 0}\), denote
\begin{equation}
    \label{eq:lattice-sum-norm}
    \norm{F}_{\Lat} := \sup_{x\in\Lat} \sum_{z\in\Lat} F\pLd{x,z}
    .
\end{equation}
Then, note that
\begin{equation}
    \norm{F_\alpha}_{\Lat}
    =
    \sup_{x\in\Lat} \sum_{r=0}^\infty
    \sumstack[r]{z\in\Lat\suchthat\\r=\Ld{x,z}} F_\alpha\pLd{x,z}
    \leq
    \Aconst \sum_{r=0}^\infty (r+1)^{\Ldim-1-\alpha}
    <
    \infty
\end{equation}
for all \(\Lambda\in \graphs\) and \(\alpha>D\) uniformly in~\(\abs{\Lambda}\).

It is important to note that all the constants will only depend on \(D\) and \(\Aconst\) and the interaction norm, even though our setting is only dealing with finite lattices \(\Lambda\in \graphs\).
In particular, given an interaction \(\Phi\) on an infinite surface regular lattice \(\Gamma\), e.g.\ \(\Gamma=\Z^D\), and defining an interaction norm on \(\Gamma\) as in~\eqref{eq:def-interaction-norm} with the sum over finite \(Z\Subset\Gamma\), the restrictions \(\Phi\rvert_{\Lambda}\) onto finite subsets \(\Lambda\subset \Gamma\) satisfy our statements uniformly in \(\Lambda\), because \(\norm{\Phi\rvert_{\Lambda}}_{\alpha,n} \leq \norm{\Phi}_{\alpha,n}\).

\section{Improved Lieb-Robinson bounds for time-dependent long-range interactions}
\label{sec:Improved-LR-bounds}

In this section we prove Lieb-Robinson bounds for long-range interacting fermions, improving and extending the previous results from~\cite{EMN2020,MKN2016}.
We begin by setting up some more notation, which will be used throughout the rest of the paper.

We say that \(H_0\colon I \rightarrow \algEven_{\Lat}\) is an on-site Hamiltonian if \(H_0(t)=\sum_{z\in\Lat} h_z(t)\) where each \(h_z\colon I \rightarrow \algEven_{\List{z}}\) is norm continuous and pointwise self-adjoint and even.

In this section, whenever there is the time-dependent interaction \(\Phi\in\interactions{\Lambda}(I)\) and an on-site Hamiltonian \(H_0\), we associate with them the operators
\begin{equation}
    H(t)
    =
    \sumstack[lr]{Z\subset\Lat} \Phi(Z,t) + H_0(t)
    \quadtext{and}
    H_{<R}(t)
    =
    \sumstack[lr]{Z\subset\Lat\suchthat\\\Ldiam{Z}<R} \Phi(Z,t) + H_0(t)
    .
\end{equation}
From continuity, it follows that there exists a unique solution \(U(t,s)\) of
\begin{equation*}
    \I \odv*{U(t,s)}{t} = H(t) \, U(t,s)
    \quadtext{with} U(s,s) = \unit
    \quadtext{for all} t,s\in I
    ,
\end{equation*}
which is a two-parameter evolution family of unitary operators, i.e.\
\begin{equation*}
    U(t,r) \, U(r,s) = U(t,s)
    \quadtext{and}
    U(t,s)^{-1} = U(t,s)^* = U(s,t)
\end{equation*}
for all \(t,r,s\in I\).
We call \(\uts{}\colon \alg_{\Lat} \rightarrow \alg_{\Lat}\) defined by
\begin{equation*}
    \uts{A} = U(t,s)^* \, A \, U(t,s) \quad\text{for all \(A\in\alg_{\Lat}\)}
\end{equation*}
the \emph{dynamics} of \(H\).
Similarly, \(\uts{}[<R]\) denotes the dynamics of \(H_{<R}\).

Extending the finite-range Lieb-Robinson bound from~\cite{MKN2016} to time-dependent Hamiltonians, we obtain the following \namecref{lem:finite-range-LR-bound-detailed}.
Within the statement, we make explicit the trivial bound \(\norm[\big]{\commutator{\uts{A}[<R], B}} \leq 2 \, \norm{A} \, \norm{B}\) which holds independently of the supports of \(A\) and \(B\).
For the convenience of the reader, the proof, a slightly stronger bound and some comments are given in \cref{app:finite-range-LR-bound-proof}.

\begin{proposition}
    \label{lem:finite-range-LR-bound}
    Let \(\Ldim\in{\N}_+\), \(\Aconst>0\), \(\alpha>\Ldim\) and \(\Lat\in\graphs\).
    Moreover, let \(I\subset{\R}\) be an interval, \(\Phi\in\interactions{\Lambda}(I)\) be a time-dependent interaction and \(H_0\) be an on-site Hamiltonian.
    For all \(X\),~\(Y\subset \Lambda\), \(A\in \alg_X\), \(B\in\alg_Y\), with \(A\) or \(B\) even, and \(R\geq1\) it holds that
    \begin{equation*}
        \norm[\big]{\commutator[\big]{\uts{A}[<R], B}}
        \leq
        \norm{A} \, \norm{B} \, \Delta\paren[\Big]{2\min\List{\abs{X}, \abs{Y}}
            \, \e^{\velocity \abs{t-s} - \Ldist{X,Y}/R}}
        ,
    \end{equation*}
    where
    \begin{equation*}
        \Delta(u) =
        \begin{cases*}
            2 & \(u>2\)   \\
            u & otherwise
        \end{cases*}
    \end{equation*}
    and \(\velocity = 2 \e^{} \, \norm{F_\alpha}_{\Lat} \, \norm{\Phi}_{\alpha}\).
\end{proposition}

It is important to note that \cref{lem:finite-range-LR-bound} is a statement about truncated long-range interactions.
This allows to have a Lieb-Robinson velocity \(\velocity\) uniform in \(R\).
For finite range interactions with uniformly bounded interaction terms, \(\sup_{Z\subset\Lat}\norm{\Phi(Z)}<J\), the velocity would increase with \(R\).
And for exponentially decaying interactions, a similar bound without dependence on~\(R\) holds~\cite{NSY2019}.
However, as can be seen by choosing large \(R\), the bound is not optimal, and we will prove better bounds for truncated long-range interactions using \cref{lem:LR-bound-iteration-step}.

The key ingredient to obtain the long-range Lieb-Robinson bounds based on \cref{lem:finite-range-LR-bound} is~\cite[Lemma~3.1]{MKN2016}, whose extension to time-dependent interactions reads as follows.
The extension to time-dependent interactions works as usual, and we only give the proof in \cref{app:splitting-LR-bound-proof} for the convenience of the reader and because it is very short with the tools presented in \cref{app:finite-range-LR-bound-proof}.

\begin{lemma}
    \label{lem:splitting-LR-bound}
    Let \(\Ldim\in{\N}_+\), \(\Aconst>0\) and \(\Lat\in\graphs\).
    Moreover, let \(I\subset{\R}\) be an interval, \(\Phi\in\interactions{\Lambda}(I)\) be a time-dependent interaction and \(H_0\) be an on-site Hamiltonian.
    Furthermore, let \(A\), \(B\in\alg_{\Lat}\), and \(1<R'<R\).
    Then,
    \begin{equation}
        \label{eq:lem-splitting-LR-bound}
        \norm[\big]{\commutator[\big]{\uts{A}[<R], B}}
        \leq
        \norm[\big]{\commutator[\big]{\uts{A}[<R'],B}}
        + 2 \, \norm{B}
        \, \sumstack[l]{Z\subset\Lat\suchthat\\\mathclap{R'\leq\Ldiam{Z}<R}}
        \, \int_{\min\List{s,t}}^{\max\List{s,t}} \norm[\big]{\commutator[\big]{\uts[t,\theta]{A}[<R'],\Phi(Z,\theta)}} \diff \theta
        .
    \end{equation}
    To obtain the bound for the complete dynamics \(\uts{}\) one can take \(R\) large enough.
\end{lemma}

The rough idea is to use the interaction picture and consider the long-range part as a perturbation of the short-range Hamiltonian.
We obtain contributions of \(A\) evolved by the short-range Hamiltonian and additional corrections due to the long-range part.
The same idea was used before in~\cite{FGC2015,MKN2016,EMN2020}, it is depicted in \cref{fig:idea-of-iteration-step1}.

\Cref{lem:splitting-LR-bound} directly allows to prove a long-range Lieb-Robinson bound as in \cref{thm:LR-bound-with-r-and-R}, by choosing \(R\) large and bounding the individual summands using the finite range bound from \cref{lem:finite-range-LR-bound}.
Instead, we first formulate the following \namecref{lem:LR-bound-iteration-step}, which allows to improve the finite-range bound and can thus be applied iteratively.
After the first application, one arrives at \cref{thm:LR-bound-with-r-and-R}, and the iterations lead to \cref{thm:LR-bound-after-iteration}.

\begin{lemma} \label{lem:LR-bound-iteration-step}
    Let \(\Ldim\in{\N}_+\), \(\Aconst>0\), \(\alpha>\Ldim\) and \(\Lat\in\graphs\).
    Moreover, let \(I\subset{\R}\) be an interval, \(\Phi\in\interactions{\Lambda}(I)\) be a time-dependent interaction and \(H_0\) be an on-site Hamiltonian.
    Fix \(s,\ t\in I\) and assume that we have a Lieb-Robinson bound
    \begin{equation}
        \label{eq:assumed-finite-range-LR-bound}
        \norm[\big]{\commutator[\big]{\uts[t,\theta]{A}[<R], B}}
        \leq
        \lambda_R\pLdist{X,Y}\min\List{\abs{X},\abs{Y}} \, \norm{A} \, \norm{B}
    \end{equation}
    for all \(X,Y\subset\Lat\), \(A\in\alg_X\), \(B\in\alg_Y\) with \(A\) or \(B\) even, and \(\theta\in\intervalcc{\min\List{s,t},\max\List{s,t}}\).
    Then, for any \(R'>1\), \eqref{eq:assumed-finite-range-LR-bound}~also holds with \(\lambda_R\) replaced by
    \begin{equation}
        \label{eq:lem-LR-bound-iteration-step}
        \tilde\lambda_R(r)
        =
        \Delta\paren[\Big]{
            2 \, \e^{\velocity \abs{t-s} - r/R'}
            + 2 \, \unit_{R>R'} \, \abs{t-s} \, \norm{\Phi}_{\alpha,1} \, F_\alpha(R') \, \norm{\lambda_{R'}}_{\Lat}
        }
        ,
    \end{equation}
    where \(\Delta\) is as in \cref{lem:finite-range-LR-bound} and \(\norm{\interpunct}_{\Lat}\) is defined in~\eqref{eq:lattice-sum-norm}.
\end{lemma}

A similar result was proven in~\cite[Theorem~2.1 and Lemma~B.1]{MKN2016} and~\cite[Lemma~1]{EMN2020}.
Our main improvement is to use the \(\norm{\interpunct}_{\alpha,1}\)-norm instead of the \(\norm{\interpunct}_{\alpha}\)-norm.
This requires a slightly stronger assumption on \(\Phi\), but it allows getting rid of a factor \(\abs{X}\) in the second summand, which was present in the previous works (where \(\lambda_R\) also had to depend on \(\abs{X}\)).
In the proof below, the difference is to use the Lieb-Robinson bound with \(\abs{Z}\) instead of \(\abs{X}\) to arrive at~\eqref{eq:proof-LR-bound-iteration-step-new}.
The improvement in particular gives a much better result in the iteratively improved Lieb-Robinson bound in \cref{thm:LR-bound-after-iteration}, where one applies \cref{lem:LR-bound-iteration-step} multiple times.
Additionally, we allow for time-dependent interactions and additional on-site Hamiltonians.

\begin{proof}[Proof of \cref{lem:LR-bound-iteration-step}]
    The surrounding \(\Delta\) is obtained from the trivial estimate for the commutator \(
        \norm{\commutator{\uts[t,\theta]{A}[<R], B}}
        \leq
        2 \, \norm{A} \, \norm{B}
    \) whenever it is better or \(X\cap Y\neq\varnothing\) and noticing that the first term in the argument of \(\Delta\) is larger than \(2\) for \(r=0\).
    It remains to concentrate on the argument of \(\Delta\).

    For \(R'\geq R\) we simply apply the Lieb-Robinson bound from \cref{lem:finite-range-LR-bound}.
    For \(R'<R\) we apply \cref{lem:splitting-LR-bound} and then bound the individual summands from~\eqref{eq:lem-splitting-LR-bound}.
    The first summand is bounded using \cref{lem:finite-range-LR-bound} by
    \begin{equation*}
        \norm[\big]{\commutator[\big]{\uts{A}[<R'], B}}
        \leq
        2 \, \e^{\velocity \, \abs{t-s} - \Ldist{X,Y}/R'} \, \abs{X} \, \norm{A} \, \norm{B}
        .
    \end{equation*}
    The second summand is bounded using the Lieb-Robinson bound from~\eqref{eq:assumed-finite-range-LR-bound} as follows.
    Note that \(\Phi\) is even, and we can thus apply the Lieb-Robinson bound~\eqref{eq:assumed-finite-range-LR-bound} for any \(A\in \alg_X\).
    To simplify notation we assume \(s<t\) and otherwise exchange their symbols.
    \begin{align*}
        \Alignindent
        2 \, \norm{B}
        \, \sumstack[l]{Z\subset\Lat\suchthat\\R'\leq\Ldiam{Z}<R}
        \int_{s}^{t}
        \norm[\big]{\commutator[\big]{\uts[t,\theta]{A}[<R'], \Phi(Z,\theta)}}
        \diff \theta
        \\&\leq
        2 \, \norm{B}
        \, \sumstack[lr]{Z\subset\Lat\suchthat\\R'\leq\Ldiam{Z}<R}
        \lambda_{R'}\pLdist{X,Z} \, \abs{Z} \, \norm{A} \int_{s}^{t} \norm{\Phi(Z,\theta)} \diff \theta
        \\&\leq
        2 \, \abs{t-s} \, \norm{A} \, \norm{B}
        \sum_{z\in\Lat} \lambda_{R'}\pLdist{X,z}
        \sup_{\theta\in\intervalcc{s,t}}\,
        \sumstack{Z\subset\Lat\suchthat\\z\in Z\\\mathclap{R'\leq\Ldiam{Z}<R}}
        \abs{Z} \, \norm{\Phi(Z,\theta)}
        \numberthis{eq:proof-LR-bound-iteration-step-new}
        \\&\leq
        2 \, \abs{t-s} \, \norm{A} \, \norm{B} \sum_{z\in\Lat} \lambda_{R'}\pLdist{X,z} \norm{\Phi}_{\alpha,1} F_\alpha(R')
        \\&\leq
        2 \, \abs{t-s} \, \norm{A} \, \norm{B} \, \norm{\Phi}_{\alpha,1} \, F_\alpha(R') \sum_{x\in X} \sum_{z\in\Lat} \lambda_{R'}\pLd{x,z}
        \\&\leq
        2 \, \abs{t-s} \, \abs{X} \, \norm{A} \, \norm{B} \, \norm{\Phi}_{\alpha,1} \, F_\alpha(R') \, \norm{\lambda_{R'}}_{\Lat}
        .
    \end{align*}
    This proves the claim with \(\abs{X}\) instead of the minimum in~\eqref{eq:assumed-finite-range-LR-bound}.
    Applying it to the RHS of
    \begin{equation*}
        \norm[\big]{\commutator[\big]{\uts{A}[<R], B}}
        = \norm[\big]{\commutator[\big]{\uts[s,t]{B}[<R], A}}
        ,
    \end{equation*}
    yields the same bound with \(\abs{Y}\), which concludes the proof.
\end{proof}

Before discussing the iteration, let us derive the bound from~\cite{MKN2016} incorporating our improvements.
To do so, we need to bound \(\norm{\lambda_{R'}}_{\Lat}\) for the finite range Lieb-Robinson bound \(\lambda_{R'}(r)=\Delta\fparen[\big]{2\exp(\velocity\abs{t-s}-r/R')}\) from \cref{lem:finite-range-LR-bound}.

Following the strategy from~\cite{MKN2016}, for any \(\rho\geq1\), we find
\begin{align*}
    \sum_{z\in\Lat} \lambda_{R'}\pLdist{x,z}
    &\leq
    \sumstack[lr]{z\in\Lat\suchthat\\\Ldist{x,z}\leq\rho} 2
    \,+
    \sum_{k=1}^\infty \sumstack[r]{z\in\Lat\suchthat\\\Ldist{x,z}=\rho+k} 2 \e^{\velocity \abs{t-s} - \Ldist{x,z}/R'}
    \\&\leq
    2 \, \Vconst \, (\rho+1)^\Ldim
    + 2 \, \e^{\velocity \abs{t-s}} \sum_{k=1}^\infty \Aconst (\rho+k)^{\Ldim-1} \e^{- (\rho+k)/R'}
    \\&\leq
    2 \, \Vconst\,(\rho+1)^\Ldim
    + 2 \, \e^{\velocity \abs{t-s}} \, \Aconst \, \int_\rho^\infty (q+1)^{\Ldim-1} \e^{- q/R'} \diff q
    \\&\leq
    2 \, \Vconst\,(\rho+1)^\Ldim
    + \LRconst \, (\rho + R')^{\Ldim-1} \,R'\, \e^{\velocity \abs{t-s} - \rho/R'}
    .
\end{align*}
Here \(\LRconst\) is a constant depending only on \(\Ldim\) and \(\Aconst\).
The bound of the integral is elementary and alternatively one can use \cref{lem:bound-integration-exp-times-polynomial}.
Now, if we choose \(R\) large enough and set \(\rho=\Ldist{X,Y}\), we obtain the first long-range Lieb-Robinson bound.%
\footnote{
    Here, and in the following the results for \(X\cap Y\neq\varnothing\), i.e.\ \(r=0\), follow directly from the trivial bound.
    And we have \(\Ldist{X,Y}\geq1\) otherwise, since we use a graph distance.
}

\begin{theorem} \label{thm:LR-bound-with-r-and-R}
    Let \(\Ldim\in{\N}_+\), \(\Aconst>0\), \(\alpha>\Ldim\) and \(\Lat\in\graphs\).
    Moreover, let \(I\subset{\R}\) be an interval, \(\Phi\in\interactions{\Lambda}(I)\) be a time-dependent interaction and \(H_0\) be an on-site Hamiltonian.
    Furthermore, let \(X,Y\subset\Lat\), \(A\in\alg_X\), \(B\in\alg_Y\), with \(A\) or \(B\) even, and \(R'\geq1\).
    Then, for any \(s,t\in I\) it holds that
    \begin{equation*}
        \norm[\big]{\commutator[\big]{\uts{A}, B}}
        \leq
        \begin{aligned}[t]
            & 2 \, \norm{A} \, \norm{B} \min\List{\abs{X},\abs{Y}}
            \\&\times
            \begin{aligned}[t]
                \paren[\Big]{
                    & \e^{\velocity \abs{t-s} - r/R'}
                    + 2 \, \Vconst \, \norm{\Phi}_{\alpha,1} \, \abs{t-s} \, (R'+1)^{-\alpha} \, (r+1)^\Ldim
                    \\& + \LRconst \, \norm{\Phi}_{\alpha,1} \, \abs{t-s} \, (R'+1)^{-\alpha} \, (r + R')^{\Ldim-1} \, R' \, \e^{\velocity \abs{t-s} - r/R'}
                },
            \end{aligned}
        \end{aligned}
    \end{equation*}
    where \(r:=\Ldist{X,Y}\), \(\velocity=2\e^{} \, \norm{F_\alpha}_{\Lat} \, \norm{\Phi}_{\alpha}\) and \(\LRconst\) is defined above.
\end{theorem}

Furthermore, choosing \(R'=r^\sigma\) with \(0<\sigma<1\) we obtain the following Lieb-Robinson bound.

\begin{corollary} \label{thm:LR-bound-with-r^1-p}
    Let \(\Ldim\in{\N}_+\), \(\Aconst>0\), \(\alpha>\Ldim\) and \(\Lat\in\graphs\).
    Moreover, let \(I\subset{\R}\) be an interval, \(\Phi\in\interactions{\Lambda}(I)\) be a time-dependent interaction, \(H_0\) be an on-site Hamiltonian, and \(\sigma\in\intervaloo{0,1}\).
    Furthermore, let \(X,Y\subset\Lat\), \(A\in\alg_X\), \(B\in\alg_Y\) with \(A\) or \(B\) even.
    Then, for any \(s,t\in I\) it holds that
    \begin{equation}
        \label{eq:lem-LR-bound-with-r^1-p}
        \begin{aligned}
            \norm[\big]{\commutator[\big]{\uts{A}, B}}
            \leq{}&
            2 \, \norm{A} \, \norm{B} \min\List{\abs{X},\abs{Y}}
            \\&\times
            \paren[\Big]{
                \e^{\velocity \abs{t-s} - r^{1-\sigma}}
                + C \, \norm{\Phi}_{\alpha,1} \, \abs{t-s} \, (r+1)^{\Ldim-\alpha\sigma} \, \paren[\big]{1+\e^{\velocity \abs{t-s} - r^{1-\sigma}}}
            }
            ,
        \end{aligned}
    \end{equation}
    where \(\velocity\) and \(r\) are the same as in \cref{thm:LR-bound-with-r-and-R} and \(C=\max\List{2\,\Vconst,2^{\Ldim(1-\sigma)}\,\LRconst}\).
\end{corollary}

The rough idea for the bound can be seen from \cref{fig:idea-of-iteration-step1}.
For the explanation it is easier to use \cref{lem:splitting-LR-bound} directly, instead of \cref{lem:LR-bound-iteration-step}.
Hence, we apply \cref{lem:splitting-LR-bound} to bound the commutator \(\norm{\commutator{\uts{A}, B}}\).
The first term in~\eqref{eq:lem-LR-bound-with-r^1-p} stems from the commutator~\(\commutator{\uts{A}[<R'],B}\) in~\eqref{eq:lem-splitting-LR-bound} after setting \(R' = r^\sigma\).
The second term stemming from the sum in~\eqref{eq:lem-splitting-LR-bound} has two contributions:
Those terms where \(Z\) intersects the light cone of \(\uts{A}[<r^\sigma]\), see~\(\Phi(Z_1)\) in \cref{fig:idea-of-iteration-step1}, are bounded using the trivial estimate.
The others, see~\(\Phi(Z_2)\) in \cref{fig:idea-of-iteration-step1}, are bounded using the finite-range Lieb-Robinson bound again and lead to the contribution containing the exponential.

\begin{figure}
    \fcapside{%
        \caption{
            Situation after applying \cref{lem:splitting-LR-bound} once, as in \cref{thm:LR-bound-with-r^1-p} or after applying \cref{lem:LR-bound-iteration-step} once.
            \newline
            The Hamiltonian is split into a short- and long-range part.
            Terms are considered short-range, if their diameter is smaller than \(R'=r^\sigma\), where \(r=\Ldist{A,B}\).
            Depicted are some terms appearing in~\eqref{eq:lem-splitting-LR-bound}:
            \(\uts<>{A}[<r^\sigma]\) is the Heisenberg time evolution of \(A\) according to the Hamiltonian \(H_{<R'}\).
            \(\Phi(Z_1)\) and \(\Phi(Z_2)\) are examples of different long-range contributions, i.e.\ \(\Ldiam{Z_1},\Ldiam{Z_2}\geq R'=r^\sigma\), appearing in~\eqref{eq:lem-splitting-LR-bound}.
        }
        \label{fig:idea-of-iteration-step1}
    }{%
        \setaltpic{1}
        \altpic{
    \tikzset{
        step1/.style={},
        step2/.style={transparent},
        step3/.style={transparent},
        step23/.style={transparent},
    }
}{
    \tikzset{
        step1/.style={transparent},
        step2/.style={},
        step3/.style={transparent},
        step23/.style={},
    }
}{
    \tikzset{
        step1/.style={transparent},
        step2/.style={transparent},
        step3/.style={},
        step23/.style={},
    }
}%
\begin{tikzpicture}[
    myLine,
    close/.style={inner sep=.1\lw},
    every node={outer sep=0pt},
]

\colorlet{Acolor}{col_1}

\node[set=Acolor, minimum width=6\ul, minimum height=5\ul, dotted, fill opacity=0.8, step1]
    (tA)        at (0,0)                                {};
\node[set=Acolor, minimum width=3.8\ul, minimum height=2.8\ul, dotted, fill opacity=0.8, step2]
    (tsA)       at (0,0)                                {};
\node[set=Acolor, minimum width=2\ul, minimum height=1\ul, label={[name=A->Z1 anchor,anchor=center]35:}]
    (A)         at (tA)                                 {\(A\)};
\node[text=Acolor, step1]
    (text tA)   at ($(tA.south)!0.5!(A.south)$)         {\(\uts<>{A}[<r^\sigma]\)};
\node[text=Acolor, step2, right,yshift=.5ex]
    (text tsA)  at (tsA.east)                          {\(\uts<>{A}[<{r'}^\sigma]\)};
\node[set=col_3, minimum width=1.5\ul, minimum height=1\ul, step1]
    (B)         at (4,-1)                               {\,\(B\)\,};

\draw[latex-latex,shorten,step1] (A) -- node[below] {\(r\)} (B);

\path let \p{A-B} = ($ (A.east) - (B.west) $),\n{dist} = {veclen(\x{A-B},\y{A-B})}
    in \pgfextra{\xdef\distAB{\n{dist}}};

\node[set=col_2, minimum width=2.8*\distAB, minimum height=2.9\ul, dotted, fill opacity=0.8, step3]
    (tsZ1)       at ($(tA.north east) + (45:0.4\ul)$)   {};
\node[text=col_2, step3, anchor=north west, yshift=.5ex]
    (text tsZ1)  at (tsZ1.south)                          {\(\uts<>{\Phi(Z_1)}[<{r'}^\sigma]\)};
\node[set=col_2, minimum width=1.7*\distAB, minimum height=1.5\ul, label={[name=Z1->A anchor,anchor=center]-155:}]
    (Z1)        at ($(tA.north east) + (45:0.4\ul)$)    {\(\Phi(Z_1)\)};
\node[set=col_3, minimum width=2*\distAB, minimum height=1.2\ul,step1, right=1\ul of B]
    (Z2)                                                 {\(\Phi(Z_2)\)\hspace*{0pt}};

\draw[latex-latex,shorten,step23] (A->Z1 anchor.center) -- node[anchor=west,yshift=-4\lw] {\(r'\)} (Z1->A anchor.center);

\path let \p{A-Z1} = ($ (A.east) - (Z1.west) $),\n{dist} = {veclen(\x{A-Z1},\y{A-Z1})}
in \pgfextra{\xdef\distAZ{\n{dist}}};

\node[set=col_3, minimum width=1.5*\distAZ, minimum height=1.1\ul,step23]
    (Z1p)       at ($(tsA.north west)+(-0.5,0.5)$)      {\(\Phi(Z_1')\)};
\node[set=col_3, minimum width=1.2*\distAZ, minimum height=1\ul,step23]
    (Z2p)       at ($(tsA.south east) + (\distAZ-0.2\ul,-0.3)$)   {\(\Phi(Z_2')\)\hspace*{0pt}};

\pgfresetboundingbox{}
\altpic{
    \node[bounding note,fit=(tA) (text tA) (Z1) (Z2) (B)] (bounding step1) {};
}{
    \node[bounding note,fit=(tsA) (text tsA) (Z1) (Z1p) (Z2p)] (bounding step2) {};
}{
    \node[bounding note,fit=(tsZ1) (text tsZ1) (Z1) (Z1p) (Z2p)] (bounding step3) {};
}

\end{tikzpicture}%
    }
\end{figure}

\begin{figure}
    \ffigbox{%
        \begin{subfloatrow}[2]
            \ffigbox{%
                \caption{
                    Second step in the iteration from~\cite{EMN2020}.
                }
                \label{fig:idea-of-iteration-step2-emn2020}
            }{%
                \setaltpic{2}
                \altpic{
    \tikzset{
        step1/.style={},
        step2/.style={transparent},
        step3/.style={transparent},
        step23/.style={transparent},
    }
}{
    \tikzset{
        step1/.style={transparent},
        step2/.style={},
        step3/.style={transparent},
        step23/.style={},
    }
}{
    \tikzset{
        step1/.style={transparent},
        step2/.style={transparent},
        step3/.style={},
        step23/.style={},
    }
}%
\begin{tikzpicture}[
    myLine,
    close/.style={inner sep=.1\lw},
    every node={outer sep=0pt},
]

\colorlet{Acolor}{col_1}

\node[set=Acolor, minimum width=6\ul, minimum height=5\ul, dotted, fill opacity=0.8, step1]
    (tA)        at (0,0)                                {};
\node[set=Acolor, minimum width=3.8\ul, minimum height=2.8\ul, dotted, fill opacity=0.8, step2]
    (tsA)       at (0,0)                                {};
\node[set=Acolor, minimum width=2\ul, minimum height=1\ul, label={[name=A->Z1 anchor,anchor=center]35:}]
    (A)         at (tA)                                 {\(A\)};
\node[text=Acolor, step1]
    (text tA)   at ($(tA.south)!0.5!(A.south)$)         {\(\uts<>{A}[<r^\sigma]\)};
\node[text=Acolor, step2, right,yshift=.5ex]
    (text tsA)  at (tsA.east)                          {\(\uts<>{A}[<{r'}^\sigma]\)};
\node[set=col_3, minimum width=1.5\ul, minimum height=1\ul, step1]
    (B)         at (4,-1)                               {\,\(B\)\,};

\draw[latex-latex,shorten,step1] (A) -- node[below] {\(r\)} (B);

\path let \p{A-B} = ($ (A.east) - (B.west) $),\n{dist} = {veclen(\x{A-B},\y{A-B})}
    in \pgfextra{\xdef\distAB{\n{dist}}};

\node[set=col_2, minimum width=2.8*\distAB, minimum height=2.9\ul, dotted, fill opacity=0.8, step3]
    (tsZ1)       at ($(tA.north east) + (45:0.4\ul)$)   {};
\node[text=col_2, step3, anchor=north west, yshift=.5ex]
    (text tsZ1)  at (tsZ1.south)                          {\(\uts<>{\Phi(Z_1)}[<{r'}^\sigma]\)};
\node[set=col_2, minimum width=1.7*\distAB, minimum height=1.5\ul, label={[name=Z1->A anchor,anchor=center]-155:}]
    (Z1)        at ($(tA.north east) + (45:0.4\ul)$)    {\(\Phi(Z_1)\)};
\node[set=col_3, minimum width=2*\distAB, minimum height=1.2\ul,step1, right=1\ul of B]
    (Z2)                                                 {\(\Phi(Z_2)\)\hspace*{0pt}};

\draw[latex-latex,shorten,step23] (A->Z1 anchor.center) -- node[anchor=west,yshift=-4\lw] {\(r'\)} (Z1->A anchor.center);

\path let \p{A-Z1} = ($ (A.east) - (Z1.west) $),\n{dist} = {veclen(\x{A-Z1},\y{A-Z1})}
in \pgfextra{\xdef\distAZ{\n{dist}}};

\node[set=col_3, minimum width=1.5*\distAZ, minimum height=1.1\ul,step23]
    (Z1p)       at ($(tsA.north west)+(-0.5,0.5)$)      {\(\Phi(Z_1')\)};
\node[set=col_3, minimum width=1.2*\distAZ, minimum height=1\ul,step23]
    (Z2p)       at ($(tsA.south east) + (\distAZ-0.2\ul,-0.3)$)   {\(\Phi(Z_2')\)\hspace*{0pt}};

\pgfresetboundingbox{}
\altpic{
    \node[bounding note,fit=(tA) (text tA) (Z1) (Z2) (B)] (bounding step1) {};
}{
    \node[bounding note,fit=(tsA) (text tsA) (Z1) (Z1p) (Z2p)] (bounding step2) {};
}{
    \node[bounding note,fit=(tsZ1) (text tsZ1) (Z1) (Z1p) (Z2p)] (bounding step3) {};
}

\end{tikzpicture}%
            }
            \ffigbox{%
                \caption{
                    Second step in our iteration after applying \cref{lem:splitting-LR-bound} with a twist.
                }
                \label{fig:idea-of-iteration-step2-our}
            }{%
                \setaltpic{3}
                \altpic{
    \tikzset{
        step1/.style={},
        step2/.style={transparent},
        step3/.style={transparent},
        step23/.style={transparent},
    }
}{
    \tikzset{
        step1/.style={transparent},
        step2/.style={},
        step3/.style={transparent},
        step23/.style={},
    }
}{
    \tikzset{
        step1/.style={transparent},
        step2/.style={transparent},
        step3/.style={},
        step23/.style={},
    }
}%
\begin{tikzpicture}[
    myLine,
    close/.style={inner sep=.1\lw},
    every node={outer sep=0pt},
]

\colorlet{Acolor}{col_1}

\node[set=Acolor, minimum width=6\ul, minimum height=5\ul, dotted, fill opacity=0.8, step1]
    (tA)        at (0,0)                                {};
\node[set=Acolor, minimum width=3.8\ul, minimum height=2.8\ul, dotted, fill opacity=0.8, step2]
    (tsA)       at (0,0)                                {};
\node[set=Acolor, minimum width=2\ul, minimum height=1\ul, label={[name=A->Z1 anchor,anchor=center]35:}]
    (A)         at (tA)                                 {\(A\)};
\node[text=Acolor, step1]
    (text tA)   at ($(tA.south)!0.5!(A.south)$)         {\(\uts<>{A}[<r^\sigma]\)};
\node[text=Acolor, step2, right,yshift=.5ex]
    (text tsA)  at (tsA.east)                          {\(\uts<>{A}[<{r'}^\sigma]\)};
\node[set=col_3, minimum width=1.5\ul, minimum height=1\ul, step1]
    (B)         at (4,-1)                               {\,\(B\)\,};

\draw[latex-latex,shorten,step1] (A) -- node[below] {\(r\)} (B);

\path let \p{A-B} = ($ (A.east) - (B.west) $),\n{dist} = {veclen(\x{A-B},\y{A-B})}
    in \pgfextra{\xdef\distAB{\n{dist}}};

\node[set=col_2, minimum width=2.8*\distAB, minimum height=2.9\ul, dotted, fill opacity=0.8, step3]
    (tsZ1)       at ($(tA.north east) + (45:0.4\ul)$)   {};
\node[text=col_2, step3, anchor=north west, yshift=.5ex]
    (text tsZ1)  at (tsZ1.south)                          {\(\uts<>{\Phi(Z_1)}[<{r'}^\sigma]\)};
\node[set=col_2, minimum width=1.7*\distAB, minimum height=1.5\ul, label={[name=Z1->A anchor,anchor=center]-155:}]
    (Z1)        at ($(tA.north east) + (45:0.4\ul)$)    {\(\Phi(Z_1)\)};
\node[set=col_3, minimum width=2*\distAB, minimum height=1.2\ul,step1, right=1\ul of B]
    (Z2)                                                 {\(\Phi(Z_2)\)\hspace*{0pt}};

\draw[latex-latex,shorten,step23] (A->Z1 anchor.center) -- node[anchor=west,yshift=-4\lw] {\(r'\)} (Z1->A anchor.center);

\path let \p{A-Z1} = ($ (A.east) - (Z1.west) $),\n{dist} = {veclen(\x{A-Z1},\y{A-Z1})}
in \pgfextra{\xdef\distAZ{\n{dist}}};

\node[set=col_3, minimum width=1.5*\distAZ, minimum height=1.1\ul,step23]
    (Z1p)       at ($(tsA.north west)+(-0.5,0.5)$)      {\(\Phi(Z_1')\)};
\node[set=col_3, minimum width=1.2*\distAZ, minimum height=1\ul,step23]
    (Z2p)       at ($(tsA.south east) + (\distAZ-0.2\ul,-0.3)$)   {\(\Phi(Z_2')\)\hspace*{0pt}};

\pgfresetboundingbox{}
\altpic{
    \node[bounding note,fit=(tA) (text tA) (Z1) (Z2) (B)] (bounding step1) {};
}{
    \node[bounding note,fit=(tsA) (text tsA) (Z1) (Z1p) (Z2p)] (bounding step2) {};
}{
    \node[bounding note,fit=(tsZ1) (text tsZ1) (Z1) (Z1p) (Z2p)] (bounding step3) {};
}

\end{tikzpicture}%
            }
        \end{subfloatrow}%
    }{%
        \caption[The pictures show the second iteration step in the proof of \cref{thm:LR-bound-after-iteration}.]{%
            Depicted is the situation after the second iteration step in the proof of \cref{thm:LR-bound-after-iteration}.
            The goal is to reduce contribution of the terms like \(\Phi(Z_1)\) in \cref{fig:idea-of-iteration-step1} which intersect the light cone of \(\uts<>{A}[<r^\sigma]\), by applying~\eqref{eq:lem-splitting-LR-bound} again.
            The situation \textcite{EMN2020} obtain is pictured in~(\subref{fig:idea-of-iteration-step2-emn2020}).
            By introducing a new sum, which scales like \(\abs{X}\), in each step, they have the bad decay in \(\abs{X}\).
            We apply~\eqref{eq:lem-splitting-LR-bound} with a twist to obtain the situation shown in~(\subref{fig:idea-of-iteration-step2-our}).
            The additional sums then scale with \(\abs{Z_1}\) which can be absorbed in the interaction norm.
            See the main text for more details.
        }
        \label{fig:idea-of-iteration-step2}
    }
\end{figure}

To further improve the dependence on \(r\), one can either apply \cref{lem:splitting-LR-bound} multiple times to the second term in~\eqref{eq:lem-splitting-LR-bound}, or apply \cref{lem:LR-bound-iteration-step} iteratively.
Both paths result in the same bound.
For the formal proof it is simpler to follow the iteration of \cref{lem:LR-bound-iteration-step}.
But here we explain the intuition using \cref{lem:splitting-LR-bound} multiple times.

As before, we split the norm using \cref{lem:splitting-LR-bound} and use the finite range Lieb-Robinson bound for the first term in~\eqref{eq:lem-splitting-LR-bound}.
The sum in the second term contains two types of terms.
The first type intersects the light cone of \(\uts{A}[<r^\sigma]\) with \(\Ldist{X,Z_1} \leq \Ldist{X,Y}\), the others lie outside \(\Ldist{X,Z_2} < \Ldist{X,Y}\), see \cref{fig:idea-of-iteration-step1}.
For the terms \(\Phi(Z)\) very close to \(X\) we use the trivial bound and the number of these terms can be bounded by \(C \, \abs{X} \, (1+\velocity \, \abs{t-s}^{\Ldim/(1-\sigma)})\).
But there are more terms, where applying \cref{lem:splitting-LR-bound} again with \(R'=\Ldist{X,Z}^\sigma\) could improve the bound, see \(\Phi(Z_1)\) in \cref{fig:idea-of-iteration-step2}.
Indeed, \cite{EMN2020} proceed by directly applying \cref{lem:splitting-LR-bound}, the resulting contributions are depicted in \cref{fig:idea-of-iteration-step2-emn2020}.
The problem is that this introduces a double sum, each contributing a factor \(\abs{X}\).
Hence, our improvement is to apply \cref{lem:splitting-LR-bound} with a twist, namely to
\begin{equation*}
    \norm[\big]{\commutator[\big]{\uts[s,t]{\Phi(Z,\theta)}[<R'], A}}
    =
    \norm[\big]{\commutator[\big]{\uts{A}[<R'], \Phi(Z,\theta)}}
    .
\end{equation*}
See the corresponding sets in \cref{fig:idea-of-iteration-step2-our}.
This sum introduces a factor \(\abs{Z_1}\) which can be absorbed in the interaction norm.
We can repeat this iteratively, e.g.\ to obtain better bounds for the contribution stemming from \(\Phi(Z'_1)\) in \cref{fig:idea-of-iteration-step2}.

\begin{theorem} \label{thm:LR-bound-after-iteration}
    Let \(\Ldim\in{\N}_+\), \(\Aconst>0\) and \(\alpha>\Ldim\).
    Then there exists a constant \(C>0\) such that for all lattices \(\Lat\in\graphs\), all \(\sigma\in\intervaloo[\big]{(\Ldim+1)/(\alpha+1),1}\), all intervals \(I\subset{\R}\), interactions \(\Phi\in\interactions{\Lambda}(I)\), on-site Hamiltonians \(H_0\), \(X\), \(Y\subset{\Lat}\), \(A\in\alg_X\), \(B\in\alg_Y\) with \(A\) or \(B\) even, and \(s,t\in I\) the following holds
    \begin{equation*}
        \norm[\big]{\commutator[\big]{\uts{A}, B}}
        \leq
        \begin{aligned}[t]
            & 2 \, \norm{A} \, \norm{B} \min\,\List{\abs{X}, \abs{Y}}
            \\&\times
            \paren[\Big]{
                \e^{\velocity \abs{t-s} - r^{1-\sigma}}
                + \sigmaconst \, (r+1)^{-\sigma\alpha} \, \velocity \abs{t-s} \, \paren[\big]{1+(\velocity \abs{t-s})^{\Ldim/(1-\sigma)}}}
            .
        \end{aligned}
    \end{equation*}
    Here \(
        \velocity=\max\,\List{2\e^{} \, \norm{F_\alpha}_{\Lat} \, \norm{\Phi}_{\alpha}, \norm{\Phi}_{\alpha,1}}
    \), \(r=\Ldist{X,Y}\) and \(
        \sigmaconst
        =
        C \, \paren[\big]{\sigma - \tfrac{D+1}{\alpha+1}}^{-2} \, \tfrac{1}{1-\sigma} \, \GammaFunc\paren[\big]{\tfrac{\Ldim}{1-\sigma}}
    \) with \(\GammaFunc\) the gamma function.
\end{theorem}

Our main improvement here compared to~\cite{EMN2020} is to obtain \(\min\,\List{\abs{X}, \abs{Y}}\) instead of a factor \(\abs{X}^n\) with \(n=\ceil{\sigma\Ldim/(\sigma\alpha-\Ldim)}+2 \rightarrow\infty\) for \(\alpha\rightarrow\Ldim\) in front of the second summand.
While this does not improve the light cone shape, it is advantageous for some applications using Lieb-Robinson bounds, e.g.\ for repetitive applications of the quasi-local inverse of the Liouvillian as it is used in the generalized adiabatic theorem, see \cref{sec:Automorphic-equivalence}.

Some readers might wonder whether the same result could be concluded using~\cite[Lemma~4]{TCE2020}, which basically allows replacing \(\abs{X}^m\) with \(\abs{X}^1\) in any existing Lieb-Robinson bound.
We explain their trick in our language using the conditional expectation in \cref{sec:comments-on-conditional-expectation} and point out why it does not work for fermions.

\section{Automorphic equivalence}
\label{sec:Automorphic-equivalence}

In this section we use the Lieb-Robinson bound from \cref{thm:LR-bound-after-iteration} to prove locality of the quasi-local inverse Liouvillian and thus locality of the spectral flow~\cite{HW2005,BMN2011,NSY2017}.
To do so, we consider a smooth family of polynomially localized Hamiltonians \(H(s)\) with a uniformly gapped part of the spectrum and show that the corresponding spectral projections \(P(s)\) are connected by a polynomially localized unitary \(P(s) = U(s)\,P(0)\,U(s)^*\).

More precisely, we require the Hamiltonians to satisfy the following assumptions additionally to certain decay properties which we impose later.

\begin{assumption}
    \label{assumption:A1}
    For a differentiable family of Hamiltonians \(s\mapsto H(s)\) with \(s\in I\) and \(I\) a closed interval, we assume that for all \(s\in I\) the spectrum \(\sigma(H(s))\) has a gapped part \(\sigma_*(s)\subset\sigma\paren{H(s)}\) satisfying the following:
    There exist \(g>0\) and continuous functions \(f_\pm\colon I\rightarrow{\R}\) such that
    \begin{thmlist}
        \item \(f_\pm(s)\in {\R}\setminus \sigma\paren{H(s)}\),
            \label{assumption:A1-f-outside-spectral-patch}
        \item \(\intervalcc[\big]{f_-(s),f_+(s)}\cap\sigma\paren{H(s)} = \sigma_*(s)\) and
        \item \(\dist[\big]{\sigma_*(s), \sigma\paren{H(s)}\setminus \sigma_*(s)}\geq g\)
    \end{thmlist}
    for all \(s\in I\).
\end{assumption}

Note that, using the standard argument involving the Cauchy formula for the resolvent, \cref{assumption:A1} implies also differentiability of \(s\mapsto P(s)\).
Taking a derivative of \(P(s)^2=P(s)\) one finds
\begin{equation*}
    \I \, \dot P(s)
    =
    \commutator[\big]{\I \commutator{\dot P(s), P(s)}, P(s)}
    =
    \commutator{\qlilgenerator_\mathrm{Kato}(s), P(s)}
    ,
\end{equation*}
with the self-adjoint Kato-generator \(\qlilgenerator_\mathrm{Kato}(s):= \I \, \commutator{\dot P(s), P(s)}\).
Hence, the solution of
\begin{equation*}
    \I \, \odv*{U_\mathrm{Kato}(s)}{s}
    =
    \qlilgenerator_\mathrm{Kato}(s) \, U_\mathrm{Kato}(s)
    \quadtext{with}
    U_\mathrm{Kato}(0)
    =
    \unit
\end{equation*}
is unitary and satisfies
\begin{equation}
    \label{eq:automorphic-equivalence}
    P(s) = U_\mathrm{Kato}(s) \, P(0) \, U_\mathrm{Kato}(s)^*
    .
\end{equation}
However, in general the Kato-generator \(\qlilgenerator_\mathrm{Kato}(s)\) does not come from a local interaction.
But \(\qlilgenerator\) is not uniquely determined by the desired property~\eqref{eq:automorphic-equivalence} and one can use the Hastings generator \(\qlilgenerator(s) = -\qliL_{H,g,0}(\dot H(s))\) instead, where \(\qliL_{H,g,\delta}\) is the \emph{quasi-local inverse of the Liouvillian} which we discuss in \cref{app:quasi-local-inverse-of-the-Liouvillian}.
In contrast to the Kato-generator, it can be given by an interaction with bounded interaction norm.
For exponentially decaying interactions this was known before~\cite{BRF2018}.
And the following \namecref{lem:quasi-local-inverse-Liouvillian}, which we prove in \cref{app:quasi-local-inverse-of-the-Liouvillian}, gives locality for polynomially decaying interactions.

\begin{proposition} \label{lem:quasi-local-inverse-Liouvillian}
    Let \(\Ldim\in{\N}_+\), \(\Aconst>0\), \(n\in{\N}\), \(\beta>0\), \(\alpha>(n+1)\,\Ldim+1+\beta\), \(g>\delta\geq0\), and \(\velocity_*\geq 1\).
    Then there exists a constant \(C>0\) such that for all lattices \(\Lat\in\graphs\) the following holds:

    Let \(\Phi\in\interactions{\Lambda}\) be an interaction such that \(
        \velocity
        =
        \max\,\List{2\e^{} \, \norm{F_\alpha}_{\Lat} \, \norm{\Phi}_{\alpha}, \norm{\Phi}_{\alpha,1}}
        \leq
        \velocity_*
    \), and let \(H_0\) be an on-site Hamiltonian.
    Then the quasi-local inverse of the Liouvillian preserves locality in the following way:
    Let \(K\) be given by an interaction \(\Phi_K\in \interactions{\Lambda}\).
    Then,
    \begin{equation*}
        \norm{\Phi_{\qliL_{H,g,\delta}(\sltoperator)}}_{\beta,n}
        \leq
        C \, \norm{\Phi_\sltoperator}_{\beta,n+1}
        .
    \end{equation*}
\end{proposition}

In contrast to the previous results, where all constants were independent of the interactions, \(C\)~includes a non-linear dependence on the norm of~\(\Phi\) through \(v_*\) and also depends on the spectral properties of the Hamiltonian~\(H\) through \(g\) and \(\delta\).
Only the linear dependence in the SLT operator~\(\sltoperator\) is specified explicitly.

Note, moreover, that there is no time-dependence in the statement.
For time-dependent Hamiltonians \(H\) and operator families \(\sltoperator\), we can define a time-dependent version
\begin{equation*}
    \paren[\big]{\qliL_{H,g,\delta}(\sltoperator)}(t)
    := \qliL_{H(t),g,\delta}\paren[\big]{\sltoperator(t)}
    ,
\end{equation*}
which does not depend on \(H\) or \(\sltoperator\) at other times.

Automorphic equivalence with an automorphism satisfying Lieb-Robinson bounds then directly follows.

\begin{theorem} \label{thm:automorphic-equivalence}
    Let \(\Ldim\in{\N}_+\), \(\Aconst>0\), \(I=\intervalcc{0,1}\), \(n\in{\N}\), \(\beta>0\), \(\alpha>(n+1)\,\Ldim+1+\beta\), \(g>\delta\geq0\), and \(\velocity_*\geq 1\).
    Then there exists a constant \(C>0\) such that for all lattices \(\Lat\in\graphs\) the following holds:

    Let \(\Phi \in \interactions{\Lambda}(I)\) be an interaction such that \(
        \velocity
        =
        \max\,\List{2\e^{} \, \norm{F_\alpha}_{\Lat} \, \norm{\Phi}_{\alpha}, \norm{\Phi}_{\alpha,1}}
        \leq
        \velocity_*
    \), and let \(H_0\) be a time-independent on-site Hamiltonian.
    Furthermore, assume that the corresponding operator family \(H\) satisfies \cref{assumption:A1}.
    Then the projections \(P(s)\) associated to \(\sigma_*(s)\) (see \cref{assumption:A1}) are unitarily equivalent
    \begin{equation}
        P(s) = U(s) \, P(0) \, U(s)^*
        ,
    \end{equation}
    where the unitaries have a local generator \(\qlilgenerator=\qliL_{H,g,0}(\dot H)\), which is given by an interaction \(\Psi\), satisfying \(\norm{\Psi(t)}_{\beta,n} \leq C \, \norm{\dot \Phi(t)}_{\beta,n+1}\).
\end{theorem}

As it was known before that the ground states could be connected by a unitary evolution, the importance of this statement lies in the fact that the generator can be given by a long-range interaction.
As such, one has locality in terms of Lieb-Robinson bounds for the unitary mapping \(A \mapsto U(s)^* \, A \, U(s)\).
To use the Lieb-Robinson bounds from \cref{sec:Improved-LR-bounds} for this purpose, one needs \(n=1\) and \(\beta>D\).
To obtain the smallest \(\alpha\), for which one has locality of the unitary mapping, one can fall back to other Lieb-Robinson.
This is reasonable here, as we are only interested in \enquote{times} up to \(s=1\) and hence do not need good light cones.
One can use~\cite{HK2006}, which only requires \(\norm{\Psi}_{\beta,1}<\infty\) for some \(\beta>0\) and then provides \(
    \norm{\commutator{U(s)^* \, A \, U(s), B}} \lesssim \norm{A} \, \norm{B} \, \abs{X} \, \abs{Y} \, F_\beta\pLdist{X,Y}
\).
Alternatively, one can use the bounds from~\cite{EMN2020,MKN2016}, which provide some slower polynomial decay in \(\Ldist{X,Y}\) without the prefactor \(\abs{Y}\) given \(\norm{\Psi}_{\beta,0}<\infty\) for some \(\beta>D\).

Similar results hold, if one starts with the previously known Lieb-Robinson bounds from~\cite{MKN2016} or~\cite{EMN2020}.
In both cases, the weaker assumption \(v = 2\e^{} \, \norm{F_\alpha}_{\Lat} \, \norm{\Phi}_{\alpha} \leq v_*\) is enough%
\footnote{%
    While~\cite{MKN2016,EMN2020} do not explicitly provide the Lieb-Robinson velocities, following the argument of \cref{lem:splitting-LR-bound}, this is the Lieb-Robinson velocity and condition in our setting.
}.
However, using the bound from~\cite{EMN2020} we would only obtain a bound \(\norm{\Psi(t)}_{\beta,n} \leq C \, \norm{\dot \Phi(t)}_{\beta,n+m}\), where \(m\) can be quite large depending on the parameters \(\alpha\) and \(\beta\), see the discussion after \cref{thm:LR-bound-after-iteration}.
And with the bound from~\cite{MKN2016}, one would only obtain \(\norm{\Psi(t)}_{\beta,n} \leq C \, \norm{\dot \Phi(t)}_{\beta,n+2}\), but additionally need to assume that \(\alpha > (n+2)\,\Ldim+1+\beta\) which in turn is a stronger assumption on~\(\Phi\).

The result obtained by using the Lieb-Robinson bound from~\cite{EMN2020} might not be a problem when dealing with \(k\)-body interactions for finite \(k\) when considering explicit Hamiltonians.
However, we want to use the results to prove generalized super-adiabatic theorems, where one iteratively applies \cref{lem:quasi-local-inverse-Liouvillian}, i.e.\ we want to apply \cref{lem:quasi-local-inverse-Liouvillian} to SLT operators \(K\) which by construction are given by many-body interactions.
There the following problem arises:
For simplicity, let us consider \(A = \qliL_{H,g,\delta}\paren[\big]{\qliL_{H,g,\delta}(\sltoperator)}\), and try to bound \(\norm{\Phi_A}_{\beta,n}\) of the corresponding interaction \(\Phi_A\).
With our bounds, we obtain
\begin{equation*}
    \norm{\Phi_A}_{\beta,n}
    \leq
    C \, \norm{\Phi_{\qliL_{H,g,\delta}(\sltoperator)}}_{\beta,n+1}
    \leq
    C \, \norm{\Phi_{\sltoperator}}_{\beta,n+2}
    ,
\end{equation*}
where, due to the second bound, \(C\) depends on \(\norm{\Phi}_{\alpha,1}\) with \(\alpha > (n+2)\,\Ldim+1+\beta\).
In comparison, with the Lieb-Robinson bounds from~\cite{EMN2020}, one would only obtain
\begin{equation*}
    \norm{\Phi_A}_{\beta,n}
    \leq
    C \, \norm{\Phi_{\qliL_{H,g,\delta}(\sltoperator)}}_{\beta,n+m}
    \leq
    C \, \norm{\Phi_{\sltoperator}}_{\beta,n+2m}
    ,
\end{equation*}
where \(C\) depends on \(\norm{\Phi}_{\alpha,0}\) with \(\alpha > (n+m+1)\,\Ldim+1+\beta\), where \(m\) can be quite large.
Hence, iteratively applied, the new bound is also significantly better concerning the scaling of the Hamiltonian.

\section{Local perturbations perturb locally}
\label{sec:LPPL}

With the results from the previous section we can also prove a \emph{local perturbation perturb locally} (LPPL) principle as introduced by \textcite{BMN2011}.
Given a time-independent interaction \(\Phi\in \interactions{\Lambda}(\intervalcc{0,1})\), an on-site Hamiltonian \(H_0\) and a perturbation \(W\in\algEven_X\) with \(X\subset\Lat\) one considers the Hamiltonian
\begin{equation}
    \label{eq:example-Hamiltonian-LPPL}
    H(s) = H_0 + \sum_{Z\subset\Lat} \Phi(Z) + s\,W
    .
\end{equation}
Supposing that \(H\) satisfies the conditions from \cref{thm:automorphic-equivalence}, we know that the projections \(P(1)\) and \(P(0)\) are unitarily equivalent.
Moreover, one can show that the generator of this unitary can be well approximated by a strictly local operator in \(\algEven_{X_R}\), such that \(P(0)\) and \(P(1)\) (almost) agree far away from \(X\).
More precisely, we find the following \namecref{cor:LPPL} where we allow more general perturbations by including them in the interaction \(\Phi\).

\begin{theorem}
    \label{cor:LPPL}
    Let \(\Ldim\in{\N}_+\), \(\Aconst>0\), \(n\in{\N}\), \(\alpha>\Ldim\), \(g>\delta\geq0\), \(\velocity_*\geq 1\) and \(\varepsilon\in \intervaloo{0,\alpha-D}\).
    Then there exists a constant \(C>0\) such that for all lattices \(\Lat\in\graphs\) the following holds:

    Let \(\Phi \in \interactions{\Lambda}(I)\) be an interaction such that \(
        \velocity
        =
        \max\,\List{2\e^{} \, \norm{F_\alpha}_{\Lat} \, \norm{\Phi}_{\alpha}, \norm{\Phi}_{\alpha,1}}
        \leq
        \velocity_*
    \), \(\dot \Phi(s) \in \algEven_X\) for some \(X\subset\Lat\) and, and let \(H_0\) be a time-independent on-site Hamiltonian.
    Furthermore, assume that the corresponding operator family \(H\) satisfies \cref{assumption:A1}.
    Then, for all \(A\in\alg_Y\) with \(Y\subset\Lat\) and \(Y\cap X = \varnothing\) it holds that
    \begin{align}
        \label{eq:cor-LPPL-trace-bound-norm}
        \abs[\big]{ \trace[\big]{P(s)\,A} -\trace[\big]{P(0)\,A} }
        &\leq
        C \, s \, \operatorname{rank}\paren[\big]{P(0)} \, \abs{Y} \, \norm{A} \, \sup_{t\in I} \, \norm{\dot H(t)} \, \paren[\big]{\Ldist{X,Y}+1}^{-\alpha+\varepsilon}
    \shortintertext{and}
        \label{eq:cor-LPPL-trace-bound-interaction-norm}
        \abs[\big]{ \trace[\big]{P(s)\,A} -\trace[\big]{P(0)\,A} }
        &\leq
        C \, s \, \operatorname{rank}\paren[\big]{P(0)} \, \abs{Y} \, \norm{A} \, \norm{\dot \Phi}_{0,1} \, \paren[\big]{\Ldist{X,Y}+1}^{-\alpha+D+\varepsilon}
        .
    \end{align}
\end{theorem}

We recall that with the prime example from~\eqref{eq:example-Hamiltonian-LPPL}, \(\sup_{t\in I} \, \norm{\dot H (t)} = \norm{W}\) and, if \(W = \sum_{Z\subset X} \Phi_W(Z)\) is given by an interaction, \(\norm{\dot \Phi}_{0,1} = \norm{\Phi_W}_{0,1}\).
While the underlying idea is similar to \cref{thm:automorphic-equivalence}, the assumptions are relaxed, because we never need to bound an interaction norm of the generator~\(G\) of~\(U\).
More precisely, for~\eqref{eq:cor-LPPL-trace-bound-norm} we consider the whole perturbation \(\dot H(t) = \sum_{Z\subset X} \dot \Phi(t)\) at once.
Hence, its operator norm is present in the bound, which is fine, if \(X\) is a fixed region.
For extensive perturbations, the bound~\eqref{eq:cor-LPPL-trace-bound-interaction-norm} is better, as it only includes an interaction norm of \(\dot \Phi\), but with a slower decay.
The proof is given in \cref{app:quasi-local-inverse-of-the-Liouvillian}.

The result is similar to the one recently obtained by \textcite{WH2022}, who prove an LPPL principle using complex analysis and avoiding the Hastings generator.
They deal with two-body interactions only, denote with \(\alphatb\) the power of the decay of each individual term, and require \(\alphatb>2\Ldim\).
Such interactions are included in our results for \(\alpha = \alphatb - \Ldim > \Ldim\), which is exactly, what we require.
However, they achieve a scaling with exponent \(\alphatb\) (minus logarithmic corrections) in~\eqref{eq:cor-LPPL-trace-bound-norm} instead of our \(\alpha-\varepsilon=\alphatb-\Ldim-\varepsilon\).
But due to the specific long-range Lieb-Robinson bound they use, they limit themselves to two-body interactions.

\section{Comments on spin systems and the conditional expectation}
\label{sec:comments-on-conditional-expectation}

In this section, we want to briefly outline, how slightly better bounds can be obtained directly from the previous bounds~\cite{MKN2016,EMN2020} for spin systems.
To do so, one uses a simple trick, which is relatively common in the quantum information community.
We formulate it here using the conditional expectation and argue, why we believe that the same argument does not work for lattice fermions.
This also poses the question, how linear light cones as in~\cite{TCE2020,TGB2021} could be obtained for lattice fermions, as the proofs rely on the same trick.

\subsection{Mathematical framework}

In this section we consider the same lattice geometries as in the rest of the work.
But instead of lattice fermions, we consider lattice spins, which requires us to change the involved Hilbert spaces and operator algebras.
On each lattice site \(z\in \Lambda\), we attach \(\spin\in \N\) spin degrees of freedom such that the local Hilbert space is \(\HS\us_z = \C^\spin\).
For any \(Z\subset \Lambda\) we then define the Hilbert space and algebra of bounded linear operators
\begin{equation*}
    \HS\us_Z
    =
    \bigotimes_{x\in Z} \HS\us_x
    \qquadtext{and}
    \alg\us_Z = \boundedOps(\HS\us_Z)
    .
\end{equation*}
For any \(Z' \subset Z\subset \Lambda\), we have \(\alg\us_{Z'}\subset \alg\us_Z\) due to the trivial identification of \(A\in \alg\us_{Z'}\) with \(A \otimes \unit_{Z\setminus Z'} \in \alg\us_{Z}\).
In comparison to the fermionic systems, for \(X\), \(Y\subset \Lambda\) disjoint, \(\commutator{A,B}=0\) for all \(A\in \alg\us_X\) and \(B\in \alg\us_Y\).
Hence, for spin systems there is no need to define an algebra of even operators and all the statements of the rest of the work hold after replacing \(\alg \to \alg\us\) and \(\algEven \to \alg\us\).

\subsection{Conditional expectation to improve Lieb-Robinson bounds in spin systems}
\label{sec:conditional-expectation-spins}

In most cases, there is no difference between the analysis of lattice spin systems or fermions on a lattice, as long as one is concerned with even operators in the latter.
In our analysis of the lattice fermions, we do restrict to even interactions and even observables to make the analysis possible and this is justified by physical reasons.

We now explain a useful trick using the conditional expectation to improve Lieb-Robinson bounds for spin systems.
In \cref{sec:conditional-expectation-fermions} we then explain, why this does not work well enough for fermions, due to the distinction between even and odd observables.
For spin systems, the conditional expectation is given by the partial trace, and can be generalized to infinite systems~\cite{NSY2019}.
It satisfies the following \namecref{lem:conditional-expectation-spins}, which is exactly what we also have for fermionic systems, but without the restriction to even operators in some of the statements, see \cref{lem:conditional-expectation}.

\begin{lemma}[{\cite[Lemma~4.1]{NSY2019}}]
    \label{lem:conditional-expectation-spins}
    Let \(X\subset\Lat\).
    Then there exists a unit-preserving, completely positive linear map \(\cexp{X}{}\colon \alg\us_{\Lat}\rightarrow\alg\us_{\Lat}\) satisfying
    \begin{thmlist}
        \item
            \(\cexp{X}{\alg\us_\Lambda}\subset\alg\us_X\);
        \item \label{lem:cmean-spins-E(ABC)=AE(B)C}
            \(\cexp{X}{ABC} = A \, \cexp{X}{B} \, C\) for all \(B\in\alg\us_{\Lat}\) and \(A,C\in\alg\us_{X}\);
            This in particular implies \(\cexp{X}{A}=A\) for all \(A\in\alg\us_{X}\);
        \item \label{lem:cmean-spins-unit-norm}
            \(\norm{\cexp{X}{}}=1\);
        \item \label{lem:cmean-spins-intersection-equals-concatenation}
            \(\cexp{X}{} \circ \cexp{Y}{} = \cexp{X\cap Y}{}\), for \(X,Y\subset\Lat\);
        \item \label{lem:cmean-spins-bound-for-difference-A-E(A)}
            If \(A\in\alg\us_{\Lat}\) satisfies
            \begin{equation}
                \norm{\commutator{A,B}}
                \leq
                \eta \, \norm{A} \, \norm{B}
                \quadtext{for all}
                B\in\alg\us_{\Lat\setminus X}
                ,
            \end{equation}
            for some \(\eta>0\), then
            \begin{equation}
                \norm{A-\cexp{X}{A}} \leq \eta \, \norm{A}
                .
            \end{equation}
    \end{thmlist}
\end{lemma}

The conditional expectation is especially useful to approximate the support of a time evolved observable, therefore let \(A\in \alg\us_{X}\).
Then, it is standard to approximate \(\uts{A}\) with the strictly local \(\cexp[\big]{X_r}{\uts{A}}\in \alg\us_{X_r}\) using \cref{lem:cmean-spins-bound-for-difference-A-E(A)} and Lieb-Robinson bound to obtain
\begin{equation*}
    \norm[\big]{
        \uts{A} - \cexp[\big]{X_r}{\uts{A}}
    }
    \leq
    \sup_{B\in \alg_{\Lambda\setminus X_r}\suchthat \norm{B}=1} \norm[\big]{\commutator[\big]{\uts{A}, B}}
    .
\end{equation*}
This indeed also works for lattice fermions and is used in the proofs.

Moreover, the conditional expectation allows for a trick which is often used for spin systems~\cite{KS2020,TCE2020,KS2021,TGB2021}.
We present it here using the conditional expectation, which we have not seen before.
Usually, the calculation is performed in an explicit realization of a partial trace, which obscures the key idea.

Let \(Y\subset \Lambda\) and choose an enumeration \(Y = \List{y_1,\dotsc,y_n}\).
Then, by \cref{lem:cmean-spins-intersection-equals-concatenation},
\begin{equation*}
    \id - \cexp{\Lambda \setminus Y}{}
    =
    \sum_{j=1}^n
    \cexp{\Lambda \setminus \List{y_1,\dotsc,y_{j-1}}}{}
    - \cexp{\Lambda \setminus \List{y_1,\dotsc,y_j}}{}
    =
    \sum_{j=1}^n
    \cexp{\Lambda \setminus \List{y_1,\dotsc,y_{j-1}}}{}
    \circ
    \paren{\id - \cexp{\Lambda \setminus \List{y_j}}{}}
    ,
\end{equation*}
where we understand \(\List{y_1,\dotsc,y_0}=\emptyset\).
And hence,
\begin{equation}
    \label{eq:decomposition-conditional-expectation}
    \norm[\big]{
        \paren{\id - \cexp{\Lambda \setminus Y}{}}(A)
    }
    \leq
    \sum_{y\in Y}
    \, \norm[\big]{
        \paren{\id - \cexp{\Lambda \setminus \List{y}}{}}(A)
    }
\end{equation}
by \cref{lem:cmean-spins-unit-norm}.
To bound the right-hand side, we can again use \cref{lem:cmean-spins-bound-for-difference-A-E(A)}, where we only need to consider operators \(B\in \alg\us_{\List{y}}\) for each summand.

Therefore, let \(f\) be an arbitrary function and assume the following Lieb-Robinson bound
\begin{equation}
    \label{eq:cmean-LRB-assumption}
    \norm[\big]{
        \commutator[\big]{\uts{A}, B}
    }
    \leq
    \norm{A} \, \norm{B}
    \, f\paren[\big]{\abs{t-s}, \Ldist{X,Y}, \min\List{\abs{X},\abs{Y}}}
\end{equation}
for all \(X\), \(Y\subset \Lambda\), \(A\in \alg\us_X\) and \(B\in \alg\us_Y\).
For \(X\), \(Y\subset \Lambda\), \(A\in \alg\us_X\) and \(B\in \alg\us_Y\) one can then bound
\begin{align}
    \nonumber
    \norm[\big]{
        \commutator[\big]{\uts{A}, B}
    }
    &\leq
    2 \, \norm{B} \, \norm[\big]{
        \paren{ \id - \cexp{\Lambda\setminus Y}{} }
        \paren[\big]{\uts{A}}
    }
    \\&\leq
    \label{eq:bound-after-first-decomposition}
    2 \, \norm{B}
    \, \sum_{y\in Y}
    \, \sup_{O_y\in \alg_{\List{y}}\suchthat \norm{O_y}=1}
    \norm[\big]{
        \commutator[\big]{\uts{A}, O_y}
    }
    \\&\leq
    \nonumber
    2 \, \norm{A} \, \norm{B}
    \, \sum_{y\in Y}
    \, f\paren[\big]{\abs{t-s}, \Ldist{X,y}, 1}
    \\&\leq
    \nonumber
    2 \, \norm{A} \, \norm{B} \, \abs{Y}
    \, f\paren[\big]{\abs{t-s}, \Ldist{X,Y}, 1}
    ,
\end{align}
where we used \(\commutator{\cexp{\Lambda\setminus Y}{\uts{A}},B} = 0\).
Applying the same argument to \(
    \norm{\commutator{\uts{A}, B}}
    =
    \norm{\commutator{A, \uts[s,t]{B}}}
\), one obtains the same result with \(\abs{Y}\) replaced by \(\abs{X}\).
Hence, under assumption~\eqref{eq:cmean-LRB-assumption}, we are able to prove the Lieb-Robinson bound
\begin{equation}
    \label{eq:cmean-LRB-first-improvement}
    \norm[\big]{
        \commutator[\big]{\uts{A}, B}
    }
    \leq
    2 \, \norm{A} \, \norm{B}
    \min\List{\abs{X},\abs{Y}} \, f\paren[\big]{\abs{t-s}, \Ldist{X,Y}, 1}
\end{equation}
for all \(X\), \(Y\subset \Lambda\), \(A\in \alg\us_X\) and \(B\in \alg\us_Y\).
For spin systems, this exactly allows changing the factor \(\abs{X}^{n}\) to \(2 \, \abs{X}\) in the Lieb-Robinson bounds from~\cite{EMN2020}, i.e.\ for spin systems one can obtain \cref{thm:LR-bound-with-r-and-R,thm:LR-bound-with-r^1-p,thm:LR-bound-after-iteration} with \(\norm{\Phi}_{\alpha,1}\) replaced by \(\norm{\Phi}_{\alpha}\) in the assumptions, after multiplying the bound with \(2\).

One can even apply the same principle to~\eqref{eq:bound-after-first-decomposition} again.
First, use \(
    \norm{\commutator{\uts{A}, O_{\List{y}}}}
    =
    \norm{\commutator{A, \uts[s,t]{O_{\List{y}}}}}
\) there, and then apply the bound~\eqref{eq:bound-after-first-decomposition} to this commutator to obtain
\begin{align}
    \nonumber
    \norm[\big]{
        \commutator[\big]{\uts{A}, B}
    }
    &\leq
    4 \, \norm{A} \, \norm{B}
    \, \sum_{y\in Y}
    \, \sup_{O_y\in \alg_{\List{y}}\suchthat \norm{O_y}=1}
    \sum_{x\in X}
    \, \sup_{O_x\in \alg_{\List{x}}\suchthat \norm{O_x}=1}
    \norm[\big]{
        \commutator[\big]{\uts{O_x}, O_y}
    }
    \\&\leq
    \label{eq:cmean-LRB-second-improvement}
    4 \, \norm{A} \, \norm{B} \, \sum_{x\in X} \, \sum_{y\in Y}
    \, f\paren[\big]{\abs{t-s}, \Ldist{x,y}, 1}
    .
\end{align}
This for example allows deducing a Lieb-Robinson bound
\begin{equation*}
    \norm[\big]{
        \commutator[\big]{\uts{A}, B}
    }
    \leq
    C \, \norm{A} \, \norm{B} \, \abs{X} \, \e^{v\abs{t-s}} \, F_{\alpha-D}\pLdist{X,Y}
\end{equation*}
for all \(X\), \(Y\subset \Lambda\), \(A\in \alg\us_X\) and \(B\in \alg\us_Y\) from the Lieb-Robinson bound from~\cite{HK2006}, which scales with \(\abs{X} \, \abs{Y}\) and thus seems useless for bounding \(\norm{\paren{\id-\cexp{X_r}{}}\uts{A}}\) using the conditional expectation at first glance.

\subsection{Difference in lattice fermions}
\label{sec:conditional-expectation-fermions}

We now explain, why the above argument does not apply to fermionic systems.
Closely comparing the statements of \cref{lem:conditional-expectation,lem:conditional-expectation-spins}, the major complication for fermions lies in the fact that the bound in \cref{lem:cmean-bound-for-difference-A-E(A)} only holds for \emph{even} \(A\) but requires a bound on \(\commutator{A,B}\) for \emph{all} \(B\in \alg_{\Lambda\setminus X}\).
Moreover, the commutator Lieb-Robinson bounds can only give a bound if at least one of the observables is even, otherwise one needs to consider anti-commutators, see~\cite{NSY2017}.
Hence, following the steps described in \cref{sec:conditional-expectation-spins}, from a Lieb-Robinson bound as in~\eqref{eq:cmean-LRB-assumption} with the restriction that \(A\) or \(B\) are even, one obtains the Lieb-Robinson bound~\eqref{eq:cmean-LRB-first-improvement} for \(A\) \emph{and} \(B\) even.
Due to the latter constraint, such a bound is not enough to characterize the localization of \(\uts{A}\) via the conditional expectation.
Moreover, we see no chance to prove a bound as in~\eqref{eq:cmean-LRB-second-improvement} for fermions, because the argument requires a Lieb-Robinson bound for arbitrary \(O_{\List{x}}\) and \(O_{\List{y}}\), which does not hold for fermions.

If the goal is to obtain a good bound on the local approximation of \(\uts{A}\), one might try to use~\eqref{eq:decomposition-conditional-expectation} directly and aim for a bound which scales with \(\abs{X}\).
Therefore, let us again assume a Lieb-Robinson bound as in~\eqref{eq:cmean-LRB-assumption} with the restriction that \(A\) or \(B\) must be even.
Then, one can bound
\begin{equation*}
    \norm[\big]{\paren[\big]{\id - \cexp{X_r}{}}\paren[\big]{\uts{A}}}
    \leq
    \sumstack[lr]{y\in \Lambda\setminus X_r}
    \, \norm[\big]{\paren[\big]{\id - \cexp{\Lambda\setminus\List{y}}{}}\paren[\big]{\uts{A}}}
    \leq
    \norm{A}
    \, \sumstack[lr]{y\in \Lambda\setminus X_r}
    \, f\paren[\big]{\abs{t-s}, \Ldist{X,y}, 1}
    .
\end{equation*}
This is in contrast to the approach for spin systems, where one first improves the Lieb-Robinson bound using the conditional expectation trick, and then directly bounds
\begin{equation*}
    \norm[\big]{\paren[\big]{\id - \cexp{X_r}{}}\paren[\big]{\uts{A}}}
    \leq
    2 \, \norm{A} \, \abs{X} \, f\paren[\big]{\abs{t-s}, \Ldist{X,Y}, 1}
    .
\end{equation*}

\begin{remark}
    Instead of localizing \(\uts{A}\) for \(A\in \alg_X\) with the conditional expectation, one can also use the truncated evolution \(\uts<X_r>{A}\in \alg_{X_r}\), generated by the interaction \(\Phi^{X_r}(Z) = \Phi(Z)\) if \(Z\subset X_r\) and \(\Phi^{X_r}(Z)=0\) otherwise.
    Similarly to the proof of \cref{lem:splitting-LR-bound}, one is then left to bound
    \begin{equation*}
        \norm[\big]{\uts{A}-\uts<X_r>{A}}
        \leq
        \sumstack[lr]{Z\subset \Lambda\suchthat\\Z\cap X_r\neq\emptyset\\Z\cap \Lambda\setminus X_r\neq\emptyset}
        \int_{\min\List{s,t}}^{\max\List{s,t}}
        \, \norm[\big]{
            \commutator[\big]{
                \Phi(Z,\theta), \uts[\theta,s]<X_r>{A}
            }
        }
        \diff \theta
        .
    \end{equation*}
    Here, one can absorb the factor \(\abs{Z}^n\) (instead of \(\abs{X}^n\)) into the interaction norm \(\norm{\Phi}_{\alpha,n}\), similarly to the proof of \cref{lem:LR-bound-iteration-step}.
    However, while the Lieb-Robinson bound one obtains this way only scales linearly in \(\abs{X}\), the assumptions on \(\Phi\) are much stronger than the ones we obtain in \cref{sec:Improved-LR-bounds}.
\end{remark}

\statement{Acknowledgments}
We thank Marius Wesle for helpful comments on an earlier version.
This work was supported by the Deutsche Forschungsgemeinschaft (DFG, German Research Foundation) through TRR~352 (470903074) and FOR~5413 (465199066).

\appendix
\crefalias{section}{appendix}
\crefalias{subsection}{appendix}

\section{Proof of the finite-range Lieb-Robinson bound}
\label{app:finite-range-LR-bound-proof}

Instead of the statement given in the main text, we will prove the following slightly stronger bound.
It is stronger for small \(\abs{t-s} \ll 1/\velocity\) and varying \(\mathord{\norm{\Phi}_{\alpha}}(r)\), which we cannot assume in the main text.

\begin{proposition}
    \label{lem:finite-range-LR-bound-detailed}
    Let \(\Ldim\in{\N}_+\), \(\Aconst>0\), \(\alpha>\Ldim\) and \(\Lat\in\graphs\).
    Moreover, let \(I\subset{\R}\) be an interval, \(\Phi\in\interactions{\Lambda}(I)\) be a time-dependent interaction and \(H_0\) be an on-site Hamiltonian.
    Furthermore, let \(X\),~\(Y\subset\Lat\) be disjoint, \(A\in\alg_X\), \(B\in\alg_Y\), with \(A\) or \(B\) even, and \(R\geq1\).
    Then, for any \(s,t\in I\)
    \begin{equation*}
        \norm[\big]{\commutator[\big]{\uts{A}[<R], B}}
        \leq
        2 \, \norm{A} \, \norm{B} \min\List{\abs{X}, \abs{Y}}
        \, \paren[\big]{\e^{I_{t,s}(\Phi)}-1} \, \e^{-\Ldist{X,Y}/R}
        ,
    \end{equation*}
    where
    \begin{equation*}
        I_{t,s}(\Phi)
        =
        2 \e^{} \, \norm{F_\alpha}_{\Lat} \int_{\min\List{t,s}}^{\max\List{t,s}}\mathord{\norm{\Phi}_{\alpha}}(\theta) \diff \theta
        \leq
        \velocity \abs{t-s}
    \end{equation*}
    and \(\velocity=2\e^{} \, \norm{F_\alpha}_{\Lat} \, \norm{\Phi}_{\alpha}\).
\end{proposition}

Combining this with the trivial bound \(\norm{\commutator{\uts{A}[<R], B}} \leq 2 \, \norm{A} \, \norm{B}\) yields \cref{lem:finite-range-LR-bound}.
In particular, for \(\Ldist{X,Y}=0\), the argument of \(\Delta\) in \cref{lem:finite-range-LR-bound} is larger than \(2\) and we use the trivial bound.

The proof of the Lieb-Robinson bound in \cref{lem:finite-range-LR-bound-detailed} uses the same strategy as the proof in~\cite{NSY2019} adjusted to our interaction norms.
Additionally, we allow time-dependent on-site contributions which do not influence the bound.
In~\cite{NSY2019} only time-independent but possibly unbounded on-site contributions were allowed by utilizing the interaction picture.
To incorporate the range \(R\) in the bound, we use the trick from~\cite{MKN2016} to see that~\eqref{eq:proof-finite-range-LR-bound-am} vanishes for \(m\,R \leq \Ldist{X,Y}\).

One might wonder why the bound does not seem to improve for larger \(\alpha\).
Indeed, it is possible to obtain an extra factor \(F_{\alpha-D}\pLdist{X,Y}\) after changing the constants.
Therefore, one adjusts the bound in~\eqref{eq:proof-finite-range-LR-bound-path} and uses the convolution property of \(F_\alpha\), see~\cite{NSY2019}.
However, since this does not improve the iteration and complicates the constants, we use the version given in \cref{lem:finite-range-LR-bound-detailed}.

Before we prove \cref{lem:finite-range-LR-bound-detailed}, let us give a lemma from~\cite{NSY2017}.
It is proven using, variation of parameters and the fact that the dynamics of a self-adjoint operator are norm preserving.

\begin{lemma}[{\cite[Lemma~3.2]{NSY2017}}]
    \label{lem:norm-estimate-variation-of-parameter}
    Let \(\HS\) be a complex Hilbert space, \(I\subset{\R}\) an interval and \(A\), \(B\colon I \rightarrow \boundedOps(\HS)\) be norm-continuous with \(A\) pointwise self-adjoint.
    Then, for any \(s\in I\), the solution of
    \begin{equation*}
        \odv*{f(t)}{t}
        = \I \, \commutator[\big]{A(t), f(t)} + B(t)
        \quadtext{with}
        f(s)=f_0\in\boundedOps(\HS)
    \end{equation*}
    satisfies
    \begin{equation*}
        \norm{f(t)}
        \leq
        \norm{f(s)}
        + \int_{\min\List{s,t}}^{\max\List{s,t}} \norm{B(\theta)} \diff \theta
        .
    \end{equation*}
\end{lemma}

Based on this, we find the following extension of~\cite[Lemma~3.3]{NSY2017}.
To state the~\namecref{lem:lemma-for-finite-range-LR-bound} we need two more definitions.
For \(k\in{\N}\) and \(X\subset\Lat\) let
\begin{equation}
    \bSets{X}
    :=
    \Set[\big]{Z\subset\Lat{} \given Z\cap X \neq \varnothing \text{ and } Z\cap\paren{\Lat \setminus X} \neq \varnothing}
\end{equation}
be the set of \emph{boundary sets} of \(X\).
And for two sets \(X,Y\subset\Lat\), let \(\delta_{X,Y}=0\) if \(X\cap Y=\varnothing\), and \(\delta_{X,Y}=1\) otherwise.

\begin{lemma}[extension of {\cite[Lemma~3.3]{NSY2017}}]
    \label{lem:lemma-for-finite-range-LR-bound}
    Let \(\Ldim\in{\N}_+\), \(\Aconst>0\) and \(\Lat\in\graphs\).
    Moreover, let \(I\subset{\R}\) be an interval, \(\Phi\in\interactions{\Lambda}(I)\) be a time-dependent interaction and \(H_0\) be an on-site Hamiltonian.
    Then, for all \(X,Y\subset\Lat\), \(A\in\alg_{X}\) and \(B\in\alg_Y\) with \(A\) or \(B\) even, it holds that
    \begin{equation*}
        \norm[\big]{\commutator[\big]{\uts{A}[<R],B}}
        \leq
        2 \, \norm{A} \, \norm{B} \, \delta_{X,Y}
        + 2 \, \norm{A} \, \sumstack[l]{\LRBoundarySet{Z}{X}\suchthat\\\Ldiam{Z}<R} \int_{\min\List{t,s}}^{\max\List{t,s}}
        \norm[\big]{\commutator[\big]{\uts{}[<R]\paren[\big]{\Phi(Z,\theta)},B}}\diff \theta
        .
    \end{equation*}
\end{lemma}

\newcommand{\calU}{\mathcal{U}}

\begin{proof}
    For better readability, we fix \(R\in{\N}\) throughout the proof and drop the subscript~‘\(_{<R}\)’.

    As before, let
    \(
        H(t) = \sum_{Z\subset\Lat\suchthat\Ldiam{Z}<R} \Phi(Z,t) + H_0(t)\), \(U(\cdot,\cdot)
    \)
    be the solution of the corresponding differential equation, and \(\uts<>{}\) be the dynamics.
    Furthermore, let
    \(
        H_X(t)
        =
        \sum_{Z\subset X\suchthat\Ldiam{Z}<R} \Phi(Z,t) + \sum_{z\in X} h_z(t)
    \), \(U_X(\cdot,\cdot)
    \)
    be the solution of the corresponding differential equation, and \(\uts<>{}[X]\) its dynamics.
    Fix \(s,t\in I\) and define
    \begin{equation*}
        f(\theta)
        =
        \commutator[\big]{\uts[\theta,s]<>{}\circ\uts[s,\theta]<>{}[X]\circ\uts<>{A}[X],B}
    \end{equation*}
    such that \(f(t)=\commutator{\uts<>{A},B}\).
    It follows that
    \begin{align*}
        \odv*{f(\theta)}{\theta}
        &=
        \I \, \commutator[\Big]{\uts[\theta,s]<>{}\paren[\Big]{\commutator[\big]{H(\theta)-H_X(\theta),\uts[s,\theta]<>{}[X]\circ\uts<>{A}[X]}}, B}
        \\&=
        \I \, \sumstack[l]{\LRBoundarySet{Z}{X}\suchthat\\\Ldiam{Z}<R} \commutator[\Big]{
            \commutator[\big]{\uts[\theta,s]<>{}\paren[\big]{\Phi(Z,\theta)},\uts[\theta,s]<>{}\circ\uts[s,\theta]<>{}[X]\circ\uts<>{A}[X]},
            B
        }
        \\&=
        \I \, \sumstack[l]{\LRBoundarySet{Z}{X}\suchthat\\\Ldiam{Z}<R} \paren[\bigg]{
            \commutator[\Big]{\uts[\theta,s]<>{}\paren[\big]{\Phi(Z,\theta)},f(\theta)}
            + \commutator[\Big]{\uts[\theta,s]<>{}\circ\uts[s,\theta]<>{}[X]\circ\uts<>{A}[X], \commutator[\big]{B, \uts[\theta,s]<>{}\paren[\big]{\Phi(Z,\theta)}}}
        }
        .
    \end{align*}
    For the second equality, we use that the commutator of all terms of \(H(\theta)-H_X(\theta)\) supported in \(\Lat\setminus X\) commute with \(\uts[s,\theta]<>{}[X]\circ\uts<>{A}[X]\subset\alg_X\) since they are even.
    This especially includes all remaining on-site terms \(h_z(\theta)\).
    For the last equality we used the Jacobi identity and the definition of \(f(\theta)\).
    Applying \cref{lem:norm-estimate-variation-of-parameter} to this equation and using the trivial bound for the outer commutator of the inhomogeneous term we find
    \begin{equation*}
        \norm[\big]{f(t)}
        \leq
        \norm[\big]{\commutator[\big]{\uts<>{A}[X],B}}
        + 2 \, \norm{A} \sumstack[l]{\LRBoundarySet{Z}{X}\suchthat\\\Ldiam{Z}<R} \int_{\min\List{s,t}}^{\max\List{s,t}} \norm[\big]{\commutator[\big]{\uts[\theta,s]<>{}\paren[\big]{\Phi(Z,\theta)},B}} \diff \theta
        .
    \end{equation*}
    The \namecref{lem:lemma-for-finite-range-LR-bound} follows using the trivial bound for the first summand.
\end{proof}

We can now give the proof of the finite-range Lieb-Robinson bound.

\begin{proof}[Proof of \cref{lem:finite-range-LR-bound-detailed}]
    For better readability, we fix \(R\in{\N}\) throughout the proof and drop the subscript \enquote{\(_{<R}\)}.
    Additionally, assume that \(s<t\) and otherwise flip the integration boundaries in the proof.

    Applying \cref{lem:lemma-for-finite-range-LR-bound} iteratively \(N\in{\N}\) times and using the trivial bound for the integrand in the last step and \(\delta_{X,Y}=0\), we find
    \begin{equation*}
        \norm[\big]{\commutator[\big]{\uts<>{A},B}}
        \leq
        2 \, \norm{A} \, \norm{B} \, \paren[\bigg]{\sum_{m=1}^N a_m(t) + R_{N+1}(t)}
    \end{equation*}
    where
    \begin{equation}
        \label{eq:proof-finite-range-LR-bound-am}
        a_m(t)
        =
        2^m
        \, \sumstack[l]{\LRBoundarySet{Z_1}{X}\suchthat\\\Ldiam{Z_1}<R}
        \, \sumstack{\LRBoundarySet{Z_2}{Z_1}\suchthat\\\Ldiam{Z_2}<R}
        \mathclap{\dotsi}
        \sumstack[r]{\LRBoundarySet{Z_m}{Z_{m-1}}\suchthat\\\Ldiam{Z_m}<R}
        \, \delta_{Y,Z_m}
        \int_s^t \int_{\mathrlap{s}}^{\mathrlap{\theta_1}} \dotsi \int_s^{\theta_{m-1}}
        \paren[\bigg]{\prod_{j=1}^m \, \norm[\big]{\Phi(Z_j,\theta_j)}}
        \diff \theta_m \dotsm \diff \theta_1
    \end{equation}
    and
    \begin{multline}
        \label{eq:proof-of-LR-bound-definition-remainder}
        R_{N+1}(t)
        =
        2^{N+1}
        \,\sumstack[l]{\LRBoundarySet{Z_1}{X}\suchthat\\\Ldiam{Z_1}<R}
        \, \sumstack[r]{\LRBoundarySet{Z_2}{Z_1}\suchthat\\\Ldiam{Z_2}<R}
        \hspacearound{1.5em}{\dotsi}
        \sumstack[l]{\LRBoundarySet{Z_{N+1}}{Z_N}\suchthat\\\Ldiam{Z_{N+1}}<R}
        \\\times
        \int_s^t \int_{\mathrlap{s}}^{\mathrlap{\theta_1}} \dotsi \int_s^{\theta_N}
        \paren[\bigg]{\prod_{j=1}^{N+1} \norm[\big]{\Phi(Z_j,\theta_j)}}
        \diff \theta_{N+1} \dotsm \diff \theta_1
        .
    \end{multline}
    We proceed with bounds for~\(a_m(t)\).
    In the same way, one shows that \({R_{N+1}(t)\rightarrow0}\) as \({N\rightarrow\infty}\).
    Noting that the sets \(X,Z_1,\dotsc,Z_{m},Y\) form a chain of pairwise intersecting sets, one can estimate
    \begin{multline}
        \label{eq:LR-bound-overcounting}
        \sumstack{\LRBoundarySet{Z_1}{X}\suchthat\\\Ldiam{Z_1}<R}
        \, \sumstack[r]{\LRBoundarySet{Z_2}{Z_1}\suchthat\\\Ldiam{Z_2}<R}
        \hspacearound{1.4em}{\dotsi}
        \sumstack[lr]{\LRBoundarySet{Z_{m}}{Z_{m-1}}\suchthat\\\Ldiam{Z_{m}}<R}
        \;\delta_{Y,Z_m}*
        \\\leq
        \sumstack[l]{w_1\in X} \sum_{w_2,\dotsc,w_{m}\in\Lat} \sumstack{w_{m+1}\in Y}
        \, \sumstack{Z_1\subset\Lat\suchthat\\w_1,w_2\in Z_1,\\\Ldiam{Z_1}<R}
        \, \sumstack[r]{Z_2\subset\Lat\suchthat\\w_2,w_3\in Z_2,\\\Ldiam{Z_2}<R}
        \hspacearound{1.3em}{\dotsi}
        \smashoperator[l]{\mathop{\sum\mathrlap{\mathop{}\;*}}_{\substack{Z_m\subset\Lat\suchthat\\w_{m},w_{m+1}\in Z_m,\\\Ldiam{Z_{m}}<R}}}
    \end{multline}
    where the \(*\) denotes arbitrary positive terms depending on all \(Z_j\).
    With that and bounding sums of products with products of sums one arrives at terms
    \begin{equation*}
        \sumstack[r]{Z_j\subset\Lat\suchthat\\w_j,w_{j+1}\in Z_j,\\\Ldiam{Z_j}<R}
        \, \norm[\big]{\Phi(Z_j,\theta_j)}
        \leq
        \norm{\Phi(\theta_j)}_{\alpha} \, F_\alpha\pLd{w_j,w_{j+1}}
        .
    \end{equation*}
    Putting everything together, we obtain
    \begin{align}
        \nonumber
        a_m(t)
        &\leq
        2^m \sumstack[l]{w_1\in X} \sum_{w_2,\dotsc,w_{m}\in\Lat} \sumstack{w_{m+1}\in Y}
        \int_s^{\mathrlap{t}} \int_{\mathrlap{s}}^{\mathrlap{\theta_1}} \dotsi \int_s^{\theta_{m-1}}
        \paren[\bigg]{\prod_{j=1}^m \, \norm[\big]{\Phi(\theta_j)}_\alpha F_\alpha\pLd{w_j,w_{j+1}}}
        \diff \theta_m \dotsm \diff \theta_1
        \\&\leq
        \label{eq:proof-finite-range-LR-bound-path}
        \min\List{\abs{X},\abs{Y}} \, \frac{1}{m!}
        \, \paren[\bigg]{
            2 \, \norm{F_\alpha}_{\Lat} \int_s^t \norm[\big]{\Phi(\theta)}_{\alpha} \diff \theta
        }^m
        .
    \end{align}
    But from~\eqref{eq:proof-finite-range-LR-bound-am} it is also apparent that \(a_m(t)=0\) for \(m\,R \leq \Ldist{X,Y}\).
    Hence, taking the limit \(N\rightarrow\infty\) we have
    \begin{align*}
        \norm[\big]{\commutator[\big]{\uts<>{A},B}}
        &\leq
        2 \, \norm{A} \, \norm{B} \min\List{\abs{X},\abs{Y}} \sumstack[l]{m>\Ldist{X,Y}/R} \frac{1}{m!}
        \, \paren[\bigg]{
            2 \, \norm{F_\alpha}_{\Lat} \int_s^t \norm[\big]{\Phi(\theta)}_{\alpha} \diff \theta
        }^m
        \e^{m-\Ldist{X,Y}/R}
        \\&\leq
        2 \, \norm{A} \, \norm{B} \min\List{\abs{X},\abs{Y}} \sumstack[l]{m\geq1} \frac{1}{m!} \paren[\bigg]{
            2 \e^{} \, \norm{F_\alpha}_{\Lat} \int_s^t \norm[\big]{\Phi(\theta)}_{\alpha} \diff \theta
        }^m
        \e^{-\Ldist{X,Y}/R}
        ,
    \end{align*}
    where \(\e^{m-\Ldist{X,Y}/R}\geq1\) was added to each summand in the first step.
\end{proof}

\section{Proof of the range-splitting lemma}
\label{app:splitting-LR-bound-proof}

The proof of \cref{lem:splitting-LR-bound} uses the techniques from \cref{app:finite-range-LR-bound-proof}.

\begin{proof}[Proof of \cref{lem:splitting-LR-bound}]
    We again fix \(R\in{\N}\) throughout the proof and drop the subscript \enquote{\(_{<R}\)}.
    Let
    \begin{equation*}
        H_{<R'}(t)
        =
        \sumstack[lr]{Z\subset\Lat\suchthat\\\Ldiam{Z}<R'\\} \Phi(Z,t) + H_0(t)
        \quadtext{and}
        H_{\geq R'}(t)
        =
        \sumstack[lr]{Z\subset\Lat\suchthat\\R'\leq\Ldiam{Z}<R\\} \Phi(Z,t)
        .
    \end{equation*}
    Furthermore, let \(U(\cdot,\cdot)\) and \(U_{<R'}(\cdot,\cdot)\) be the solution of the differential equations corresponding to \(H(t)=H_{<R'}(t)+H_{\geq R'}(t)\) and \(H_{<R'}\), respectively, and \(\uts<>{}\) and \(\uts<>{}[<R']\) their dynamics.
    Let \(\calU(t,s)=U_{<R'}(t,s)^*\,U(t,s)\).
    Fix \(s,t\in I\) and define
    \begin{equation*}
        f(\theta)
        =
        \commutator[\big]{\uts<>{A}[<R'],\calU(\theta,s)\, B \, \calU(\theta,s)^*}
        .
    \end{equation*}
    Then
    \(
        \norm{f(t)}
        = \norm{\commutator{\uts<>{A},B}}
        .
    \)
    Furthermore, from
    \begin{equation*}
        \odv*{f(\theta)}{\theta}
        =
        \begin{aligned}[t]
            &\commutator[\big]{-\I \, \uts[\theta,s]<>{}\paren[\big]{H_{\geq R'}(\theta)},f(\theta)}
            \\& + \I \commutator[\Big]{\calU(\theta,s)\,B\,\calU(\theta,s)^*,\commutator[\big]{\uts<>{A}[<R'],\uts[\theta,s]<>{}[<R']\paren[\big]{H_{\geq R'}(\theta)}}}
        \end{aligned}
    \end{equation*}
    it follows with \cref{lem:norm-estimate-variation-of-parameter} that
    \begin{align*}
        \norm[\big]{\commutator[\big]{\uts<>{A},B}}
        &\leq
        \norm[\big]{\commutator[\big]{\uts<>{A}[<R'],B}}
        + 2 \, \norm{B} \int_{\min\List{s,t}}^{\max\List{s,t}} \norm[\big]{\commutator[\big]{\uts<>{A}[<R'],\uts[\theta,s]<>{H_{\geq R'}(\theta)}[<R']}} \diff \theta
        \\&\leq
        \norm[\big]{\commutator[\big]{\uts<>{A}[<R'],B}}
        + 2 \, \norm{B} \sumstack[l]{Z\subset\Lat\suchthat\\\mathclap{R'\leq\Ldiam{Z}<R}} \int_{\min\List{s,t}}^{\max\List{s,t}} \norm[\big]{\commutator[\big]{\uts[t,\theta]<>{A}[<R'],\Phi(Z,\theta)}} \diff \theta
        .
    \end{align*}
\end{proof}

\section{Proof of the iteratively improved Lieb-Robinson bound}
\label{sec:proof-iteratively-improved-LR-bound}

For the proof of \cref{thm:LR-bound-after-iteration}, we follow the strategy from~\cite{EMN2020} and iteratively apply \cref{lem:LR-bound-iteration-step} to obtain an improved Lieb-Robinson bound.
We repeat the proof here to achieve the better dependence on \(\abs{X}\), adjust the proof to our definition of the graphs and metrics and find the correct scaling of the constant depending on \(\sigma\).

The lattice parameters \(\Aconst\) and \(D\) as well as the decay exponent \(\alpha\) are fixed in the beginning, and all the constants might depend on them.
Importantly, the constants do not depend on the specific lattice \(\Lambda\in \graphs\), the interaction or the parameter~\(\sigma\).

\begin{proof}[Proof of \cref{thm:LR-bound-after-iteration}]
    First notice that the \namecref{thm:LR-bound-after-iteration} is trivially satisfied for \(r=0\) by using the trivial bound.
    Hence, we assume \(r\geq1\) in the following.

    The following \namecref{lem:bound-lattice-sum-by-integral} is similar to~\cite[Lemma~2]{EMN2020} adopted to our geometries.
    It allows bounding \(\norm{\lambda}_{\Lat}\) by an integral as
    \begin{equation}
        \label{eq:example-bound-of-LRsum}
        \norm{\lambda}_{\Lat}
        \leq
        \lambda(0) + \latticesumconst \int_{1/2}^{\infty} \lambda(\rho) \, \rho^{\Ldim-1} \diff \rho
        .
    \end{equation}

    \begin{lemma}\label{lem:bound-lattice-sum-by-integral}
        Let \(f\colon {\R}\rightarrow{\R}\) be a monotonically decreasing function, \(\Aconst>0\), \(\Ldim\in{\N}\), \(\Lambda\in\graphs\), \(x\in\Lat\) and \(R\geq1\).
        Then, it holds that
        \begin{equation*}
            \sumstack[r]{z\in\Lat\suchthat\\1\leq\Ld{x,z}\leq R} f\pLd{x,z}
            \leq
            \latticesumconst \int_{1/2}^R f(r) \, r^{\Ldim-1} \diff r
            ,
        \end{equation*}
        with \(\latticesumconst = 2^\Ldim \, \Aconst \geq 1\).
    \end{lemma}

    \begin{proof}
        Since \(\Ld{}\) is integer valued and \(f\) is monotonically decreasing we find
        \begin{align*}
            \sumstack[r]{z\in\Lat\suchthat\\1\leq\Ld{x,z}\leq R} f\pLd{x,z}
            &=
            \sum_{j=1}^R
            \sumstack[r]{z\in\Lat\suchthat\\\Ld{x,z}=j} f\pLd{x,z}
            \\&\leq
            \Aconst
            \sum_{j=1}^R
            j^{\Ldim-1} \, f(j)
            \\&\leq
            2\,\Aconst
            \int_{1/2}^{R} f(r) \, \paren{r+1/2}^{\Ldim-1} \diff r
            \\&\leq
            2^\Ldim\,\Aconst
            \int_{1/2}^{R} f(r) \, r^{\Ldim-1} \diff r
            .
            \qedhere
        \end{align*}
    \end{proof}

    To bound the integral in~\eqref{eq:example-bound-of-LRsum}, we also need the following \namecref{lem:bound-integration-exp-times-polynomial} from~\cite{EMN2020}.
    Since we want to provide explicit scaling, we state the result with an explicit constant that can easily be deduced from the original proof in~\cite{EMN2020}.

    \begin{lemma}[{\cite[Lemma~3]{EMN2020}}]
        \label{lem:bound-integration-exp-times-polynomial}
        For \(\mu\in{\R}\) and \(\nu>0\) and
        \begin{equation*}
            \INTconst
            =
            \tfrac{1}{\nu} \, \max \, \Set[\Big]{1,\e^{}\,\GammaFunc\paren[\Big]{\tfrac{\mu+1}{\nu}}}
            ,
        \end{equation*}
        where \(\GammaFunc\) is the gamma function,
        it holds that
        \begin{equation*}
            \int_\rho^\infty \e^{-x^\nu} x^\mu \diff x
            \leq
            \INTconst \, \e^{-\rho^\nu}\paren[\big]{1+\rho^{\mu-\nu+1}}
            \quadtext{for all \(\rho>0\).}
        \end{equation*}

    \end{lemma}

    Now fix \(t\) and \(s\), and write \(\delta=\abs{t-s}\).
    \newcommand\vd{\velocity \delta}
    We begin with the finite range Lieb-Robinson bound from \cref{lem:finite-range-LR-bound}, \(\lambda^{(0)}_{R'}(r) = \Delta(2\,\e^{\velocity \delta-r/{R'}})\).
    The trivial bound is in particular better for \(r\leq R'\vd\).
    Hence, for \(R'\vd>1/2\), we find
    \begin{align*}
        \norm[\big]{\lambda^{(0)}_{R'}}_{\Lat}
        &\leq
        2
        + 2 \, \latticesumconst \, \paren[\Bigg]{
            \int_{1/2}^{R'\vd} \rho^{\Ldim-1} \diff \rho
            + \int_{R'\vd}^\infty \e^{\vd -\rho/R'} \, \rho^{\Ldim-1} \diff \rho
        }
        \\&\leq
        2 \, \latticesumconst \, \paren[\Big]{
            1
            + (R'\vd)^\Ldim
            + \INTconst[\Ldim-1,1] \, \e^{\vd} R'^\Ldim \, \e^{-\vd} \, \paren[\big]{1+(\vd)^{\Ldim-1}}
        }
        \\&\leq
        6 \, \latticesumconst \, \INTconst[\Ldim-1,1] \, R'^\Ldim \, \paren[\big]{1+(\vd)^\Ldim}
    \end{align*}
    where we substituted \(\rho/R'\rightarrow\rho\) and used \cref{lem:bound-integration-exp-times-polynomial} in the second step.
    For the last step, we restricted to \(R'\geq1\).
    For \(R'\vd\leq1/2\), we do not split the integral and instead obtain
    \begin{align*}
        \norm[\big]{\lambda^{(0)}_{R'}}_{\Lat}
        &\leq
        2
        + 2 \, \latticesumconst
        \, \int_{1/2}^\infty \e^{\vd -\rho/R'} \, \rho^{\Ldim-1} \diff \rho
        \\&\leq
        2 \, \latticesumconst \, \paren[\bigg]{
            1
            + \INTconst[\Ldim-1,1] \, \e^{\vd} R'^\Ldim \, \e^{-\tfrac{1}{2R'}} \, \paren[\Big]{1+\paren[\big]{\tfrac{1}{2R'}}^{\Ldim-1}}
        }
        \\&\leq
        6 \, \latticesumconst \, \INTconst[\Ldim-1,1] \, R'^\Ldim \, \paren[\big]{1+(\vd)^\Ldim}
        .
    \end{align*}
    In the last step, we added the term \((\vd)^\Ldim\) to obtain the same bound as for \(R'\vd>1/2\).

    Setting \(R'=r^\sigma\), \cref{lem:LR-bound-iteration-step} yields the improved Lieb-Robinson bound
    \begin{equation*}
        \lambda^{(1)}_R(r)
        \leq
        \Delta\paren[\Big]{
            2 \, \e^{\vd-r^{1-\sigma}} + 2 \, C_1 \, \unit_{R>r^\sigma} \, \delta \, \norm{\Phi}_{\alpha,1} \, (r+1)^{-\sigma(\alpha-\Ldim)} \, \paren[\big]{1+(\vd)^\Ldim}
        }
        ,
    \end{equation*}
    with \(
        C_1
        =
        2^{\alpha(1-\sigma)+1} \, 3 \, \latticesumconst \, \INTconst[\Ldim-1,1]
    \).
    The additional prefactor comes from bounding \(F_\alpha(r^\sigma) \leq 2^{\alpha(1-\sigma)} \, (1+r)^{\sigma\alpha}\).

    We now proceed by induction.
    Assume that we have a Lieb-Robinson bound with
    \begin{equation}
        \label{eq:LR-bound-in-iteration}
        \lambda^{(n)}_R(r)
        \leq
        \Delta\paren[\Big]{
            2 \, \e^{\vd-r^{1-\sigma}} + 2 \, \unit_{R>r^\sigma} \, \sum_{i=1,2} f^{(n)}_i(\vd) \, {(r+1)}^{\mu^{(n)}_i}
        }
        .
    \end{equation}
    For \(n=1\) this is satisfied for
    \begin{align*}
        f^{(1)}_1(\tau) &= C_1 \, (\tau+\tau^{\Ldim+1}),
        & \mu^{(1)}_1 &= \sigma \, (-\alpha+\Ldim),
        \\
        f^{(1)}_2(\tau) &= 0 \quad\text{and}
        & \mu^{(1)}_2 &= -\sigma\alpha
        ,
    \end{align*}
    when we redefine \(
        \velocity=\max\List{2\e^{} \, \norm{F_\alpha}_{\Lat} \, \norm{\Phi}_{\alpha}, \norm{\Phi}_{\alpha,1}}
    \) to keep the constant independent of~\(\Phi\).

    We now calculate \(\norm[\big]{\lambda^{(n)}_{R'}}_{\Lat}\).
    Only looking at the exponential term in~\eqref{eq:LR-bound-in-iteration}, we notice that the trivial bound is better at least for \(r < (\vd)^{1/(1-\sigma)}\).
    We first consider the case \(\vd \geq 1\).
    Additionally, we assume \(\Ldim+\mu^{(n)}_i \neq 0\) which we ensure later by changing the iteration at the right step.
    Then
    \begin{align*}
        \norm[\big]{\lambda^{(n)}_{R'}}_{\Lat}
        &\leq
        2
        + 2 \, \latticesumconst
        \, \int_{1/2}^{(\vd)^{1/(1-\sigma)}} \rho^{\Ldim-1} \diff \rho
        + \latticesumconst
        \, \int_{(\vd)^{1/(1-\sigma)}}^\infty \rho^{\Ldim-1} \, \lambda^{(n)}_{R'}(\rho) \diff \rho
        \numberthis{eq:proof-iterative-LR-bound-bound-LRint}
        \\&\leq
        2 \, \latticesumconst \,
        \begin{aligned}[t]
            \paren[\Bigg]{
                1
                + {(\vd)}^{\frac{\Ldim}{1-\sigma}}
                &+ \int_{(\vd)^{1/(1-\sigma)}}^\infty \rho^{\Ldim-1} \, \e^{\vd-\rho^{1-\sigma}} \diff \rho
                \\&+ \sumstack[l]{i=1,2} \int_{(\vd)^{1/(1-\sigma)}}^{R'^{1/\sigma}} f^{(n)}_{i}(\vd) \, (\rho+1)^{\Ldim-1+\mu^{(n)}_i} \diff \rho
            }
        \end{aligned}
        \numberthis{eq:proof-iterative-LR-bound-LRint-rho-part}
        \\&\leq
        2 \, \latticesumconst \,
        \begin{aligned}[t]
            \paren[\Bigg]{
                1
                + {(\vd)}^{\frac{\Ldim}{1-\sigma}}
                &+ \INTconst[\Ldim-1,1-\sigma] \, \paren[\Big]{1+{(\vd)}^{\frac{\Ldim}{1-\sigma}-1}}
                \\&+ \sumstack[l]{i=1,2} \frac{f^{(n)}_{i}(\vd)}{\Ldim+\mu^{(n)}_i} \, {(\rho+1)}^{\Ldim+\mu^{(n)}_i} \biggr\rvert_{(\vd)^{1/(1-\sigma)}}^{R'^{1/\sigma}}
            }
        \end{aligned}
        \\&\leq
        \begin{multlined}[t]
            4 \, \latticesumconst \, \INTconst[\Ldim-1,1-\sigma]
            \, \paren[\Big]{1+{(\vd)}^{\frac{\Ldim}{1-\sigma}}}
            \\+ 2 \, \latticesumconst \, \sumstack[l]{i=1,2} \frac{f^{(n)}_{i}(\vd)}{\abs[\big]{\Ldim+\mu^{(n)}_i}}
            \begin{cases}
                \paren[\big]{R'^{1/\sigma}+1}^{\Ldim+\mu^{(n)}_i}
                & \Ldim+\mu^{(n)}_i > 0
                \\
                (\vd)^{(\Ldim+\mu^{(n)}_i)/(1-\sigma)}
                & \Ldim+\mu^{(n)}_i < 0.
            \end{cases}
        \end{multlined}
    \end{align*}
    In the last step, we bounded the last summand by the upper limit of the integration if \(\Ldim+\mu^{(n)}_i>0\) and by the lower if \(\Ldim+\mu^{(n)}_i<0\) because the other limit gives a negative contribution in each case.
    In the latter case we moreover removed the “\({}+1\)” in the parenthesis.

    Using \cref{lem:LR-bound-iteration-step} and setting \(R'=r^\sigma\), we obtain \(\lambda^{(n+1)}_R\) as in~\eqref{eq:LR-bound-in-iteration} with
    \begin{align*}
        f^{(n+1)}_1(\tau)
        &=
        \tfrac{C_0}{\abs[\big]{D+\mu^{(n)}_1}} \, \tau \, f^{(n)}_1(\tau),
        & \mu^{(n+1)}_1
        &=
        -\sigma\,\alpha + \Ldim + \mu^{(n)}_1
        & \text{if \(\Ldim+\mu^{(n)}_1>0\),}
        \\
        f^{(n+1)}_1(\tau)
        &=
        \tfrac{C_0}{\abs[\big]{D+\mu^{(n)}_1}} \, \tau^{1+(\Ldim+\mu^{(n)}_1)/(1-\sigma)} f^{(n)}_1(\tau),
        & \mu^{(n+1)}_1
        &=
        -\sigma\alpha
        & \text{if \(\Ldim+\mu^{(n)}_1<0\),}
    \intertext{and}
        f^{(n+1)}_2(\tau)
        &=
        \mathrlap{
            C_2 \, \paren[\big]{\tau+\tau^{1+\Ldim/(1-\sigma)}}
            + \tfrac{C_0}{\abs{D-\sigma\alpha}} \, \tau^{1+(\Ldim-\sigma\alpha)/(1-\sigma)} f^{(n)}_2(\tau),
            \qquad
            \mu^{(n+1)}_2 = -\sigma\alpha.
        }
    \end{align*}
    The constants are given by \(C_0 = 2^{\alpha(1-\sigma)+1} \, \latticesumconst\) and \(C_2 = 2^{\alpha(1-\sigma)+2} \, \latticesumconst \, \INTconst[\Ldim-1,1-\sigma]\) and for \(f^{(n)}_2\) we already used that \(\mu^{(n)}_2=-\sigma\alpha\) is constant.

    Let \(\varepsilon=\sigma\alpha-D>0\), then the iteration directly yields
    \begin{equation*}
        f^{(n)}_2(\tau)
        =
        C_2 \, (\tau+\tau^{1+\Ldim/(1-\sigma)}) \sum_{j=0}^{n-2} \paren[\Big]{\tfrac{C_0}{\varepsilon}}^j \tau^{-j\frac{\sigma(\alpha+1)-(\Ldim+1)}{1-\sigma}}
        ,
    \end{equation*}
    so that we can expect a polynomial scaling in \(\tau\) with exponent at least \(1+D/(1-\sigma)\).

    Now assume that \(D+\mu^{(1)}_1>0\), then
    \begin{equation*}
        \mu^{(n)}_1
        =
        -n\sigma\alpha + (n-1+\sigma)\,\Ldim
    \end{equation*}
    decreases by \(\varepsilon\) in each iteration step until \(\mu^{(n)}_1-D \leq 0\).
    To avoid small \(\abs[\big]{D+\mu^{(n)}_1}\), which would give large multiplicative constants, we tweak the iteration as follows:
    First choose \(\eta \in \intervaloo{0,1/2}\) and
    \begin{equation*}
        n_*
        =
        \ceil*{\frac{\sigma\Ldim}{\sigma\alpha-\Ldim}-(1-\eta)}
        <
        \frac{\sigma\Ldim}{\sigma\alpha-\Ldim}+\eta
    \end{equation*}
    so that the basic iteration from above gives \(
        \Ldim+\mu_1^{(n_*)}\in\intervaloc[\big]{-\varepsilon\,\eta, \varepsilon\,(1-\eta)}
    \).

    If \(
        \Ldim+\mu_1^{(n_*)}\in\intervaloc[\big]{\varepsilon\,\eta, \varepsilon\,(1-\eta)}
    \), we just continue the iteration and obtain \(\Ldim+\mu_1^{(n_*+1)}=-\varepsilon\,\eta<0\) and
    \begin{equation*}
        f^{(n_*+2)}_1(\tau)
        \leq
        \paren[\Big]{\tfrac{C_0}{\varepsilon}}^{n_*+1} \, \tfrac{1}{\eta^2} \, \tau^{1+\frac{D}{1-\sigma}} \, C_1 \, \paren[\big]{\tau^{-\Ldim}+1}
        ,
    \end{equation*}
    because
    \begin{equation*}
        1+\frac{\Ldim+\mu^{(n_*+1)}_1}{1-\sigma} + (n_*+1) + D
        =
        1+\frac{D}{1-\sigma} - \frac{n_*+1}{\sigma-1}\paren[\big]{\sigma\,(\alpha+1)-(D+1)}
        \leq
        1 + \frac{D}{1-\sigma}
        .
    \end{equation*}
    Thus, the scaling in \(\tau\) is already better than that of \(f_2^{(n)}\) and also the spatial decay does not improve in further steps.

    Otherwise, if the basic iteration yields \(\Ldim+\mu_1^{(n_*)}\in\intervaloc[\big]{-\varepsilon\,\eta, \varepsilon\,\eta}\), we instead choose \(\mu^{(n_*)}_1=-D+\varepsilon\,\eta\) by bounding the term in~\eqref{eq:LR-bound-in-iteration}.
    Then, \(\Ldim+\mu_1^{(n_*+1)}=-\varepsilon\,(1-\eta) < 0\).
    And the iteration results in
    \begin{equation*}
        f^{(n_*+2)}_1(\tau)
        \leq
        \paren[\Big]{\tfrac{C_0}{\varepsilon}}^{n_*+1} \, \tfrac{4}{\eta} \, \tau^{1+\frac{D}{1-\sigma}+\eta\,\paren[\big]{1+\frac{\sigma\,\alpha-D}{1-\sigma}}} \, C_1 \, \paren[\big]{\tau^{-\Ldim}+1}
        .
    \end{equation*}
    After a further iteration step and choosing \(\eta = \sigma - \frac{D+1}{\alpha+1}\), we obtain
    \begin{equation*}
        f^{(n_*+3)}_1(\tau)
        \leq
        \paren[\Big]{\tfrac{C_0}{\varepsilon}}^{n_*+2} \, \tfrac{4}{\eta} \, \tau^{1+\frac{D}{1-\sigma}} \, C_1 \, \paren[\big]{\tau^{-\Ldim}+1}
        ,
    \end{equation*}
    which again has a better scaling in \(\tau\) than \(f_2^{(n)}\).

    Since we can bound
    \begin{equation*}
        \varepsilon > \tfrac{\alpha-\Ldim}{\alpha+1} > 0
        \qquadtext{and}
        n_* < \tfrac{2\,\alpha\,D}{\alpha-D} + \tfrac{3}{2}
        ,
    \end{equation*}
    the only \(\sigma\)-dependence of the constants left is in
    \begin{equation*}
        \INTconst[\Ldim-1,1-\sigma]
        <
        \tfrac{1}{1-\sigma} \, \GammaFunc\paren[\Big]{\tfrac{\Ldim}{1-\sigma}}
        \qquadtext{and}
        \eta
        =
        \sigma - \tfrac{D+1}{\alpha+1}
    \end{equation*}
    and leads to divergences at the boundaries of the interval of allowed \(\sigma\).
    This completes the first part of the proof.

    We are left with the case \(\vd<1\), where we do not split the integral in~\eqref{eq:proof-iterative-LR-bound-bound-LRint} and instead obtain
    \begin{equation*}
        \LRint[\lambda^{(n)}_{R'}]
        \leq
        16 \, \latticesumconst \, \INTconst[\Ldim-1,1-\sigma]
        + 2 \, \latticesumconst \, \sumstack[l]{i=1,2} \frac{f^{(n)}_{i}(\vd)}{\abs[\big]{\Ldim+\mu^{(n)}_i}}
        \begin{cases}
            \paren[\big]{R'^{1/\sigma}+1}^{\Ldim+\mu^{(n)}_i}
            & \Ldim+\mu^{(n)}_i > 0
            \\
            \paren[\Big]{\frac{3}{2}}^{\Ldim+\mu^{(n)}_i}
            & \Ldim+\mu^{(n)}_i < 0.
        \end{cases}
    \end{equation*}
    This results in
    \begin{align*}
        f^{(n+1)}_1(\tau)
        &=
        \tfrac{C_0}{\abs[\big]{D+\mu^{(n)}_1}} \, \tau \, f^{(n)}_1(\tau),
        &&
        \begin{aligned}
            \mu^{(n+1)}_1
            &=
            -\sigma\,\alpha + \Ldim + \mu^{(n)}_1
            &&\text{if \(\Ldim+\mu^{(n)}_1 > 0\),}\\
            \mu^{(n+1)}_1
            &=
            -\sigma\,\alpha
            &&\text{if \(\Ldim+\mu^{(n)}_1 < 0\),}
        \end{aligned}
    \shortintertext{and}
        f^{(n+1)}_2(\tau)
        &=
        4 \, C_2 \, \tau + \tfrac{C_0}{\abs{D-\sigma\alpha}} \, \tau \, f^{(n)}_2(\tau),
        && \mu^{(n+1)}_2
        =
        -\sigma\alpha
    \end{align*}
    with the same adjustment of the \(n_*\) step as before,
    and yields
    \begin{equation*}
        f^{(n)}_1(\tau)
        =
        \paren[\Big]{\tfrac{C_0}{\varepsilon}}^{n-1} \, \tfrac{2}{\eta^2} \, C_1 \, \tau^n \, \paren[\big]{1+\tau^\Ldim}
        ,\qquad
        f^{(n)}_2(\tau)
        =
        4 \, C_2 \sum_{j=1}^{n-1} \paren[\Big]{\tfrac{C_0}{\varepsilon}}^{j} \, \tau^j
    \end{equation*}
    and \(\mu_1^{(n)} = \mu_2^{(n)} = -\sigma\alpha\) for \(n\geq n_*+2\).
    We again stop at \(n=n_*+2\) or \(n=n_*+3\) and the lowest power in \(\tau\) (recall that \(\tau<1\) here) is \(1\).
    Putting everything together proves the \namecref{thm:LR-bound-after-iteration}.
\end{proof}

\section{The spectral flow and its decay properties}
\label{app:quasi-local-inverse-of-the-Liouvillian}

In this section we define the spectral flow \(\qliL_{H,g,\delta}\) and prove its decay properties as stated in \cref{lem:quasi-local-inverse-Liouvillian}.
Therefore, let \(g>\delta\geq0\) and \(\Wfgd\in L^1({\R})\) be a function satisfying
\begin{equation*}
    \sup_{\abs{s}>1}\,\abs{s}^n \, \abs{\Wfgd(s)} < \infty
    \quad\text{for all \(n\in{\N}\)}
\end{equation*}
with Fourier transform \(\ftWf_{g,\delta}\in C^\infty({\R})\) satisfying
\begin{equation*}
    \ftWf_{g,\delta}(\omega) = \frac{-\I}{\sqrt{2\pi}\,\omega}
    \quad\text{for all \(\abs{\omega}\geq g\)}
    \qquadtext{and}
    \ftWf_{g,\delta}(\omega) = 0
    \quad\text{for all \(\abs{\omega}\leq\delta\)}
    .
\end{equation*}
For \(\delta=0\) such a function was constructed in~\cite{BMN2011}, it additionally satisfies \(\norm{\Wf_{g,0}}_{L^1} = 1\) and \(\norm{\Wf_{g,0}}_{\infty} = 1/2\), and is enough to prove automorphic equivalence.
This function was then modified in~\cite{MT2019,Teufel2020} for \(\delta>0\),
because the spectral flow has additional properties for larger \(\delta\), see \cref{prop:qliL-general-properties}.

On a general Hilbert space we now find the following \namecref{prop:qliL-general-properties}.
The proof and more details are given in~\cite{MT2019,Teufel2020}.
The general idea goes back to the construction of the spectral flow in~\cite{Hastings2004,HW2005}, see also~\cite{BMN2011,BRF2018}.

For any self-adjoint Hamiltonian \(H\in \algEven_\Lambda\), we can define
\begin{equation}
    \label{eq:definition-qliL}
    \qliL_{H,g,\delta}\colon\alg_{\Lat}\rightarrow\alg_{\Lat},
    \quad
    \qliL_{H,g,\delta}(A)
    =
    \int_{\R} \Wf_{g,\delta}(s) \, \e^{\I H s} \, A \, \e^{-\I H s} \diff s
\end{equation}
which by the properties of \(\Wf\) satisfies the following.

\begin{proposition}
    \label{prop:qliL-general-properties}
    Let \(H\in \algEven_\Lambda\) be self-adjoint and assume that the spectrum \(\sigma(H)\) has a gapped part \(\sigma_*\subset\sigma\paren{H}\) such that \(\sigma_*\subset I\), \(\sigma(H)\setminus \sigma_* \subset \R\setminus I\) and \(\dist[\ig]{\sigma_*, \sigma\paren{H}\setminus \sigma_*}\geq g\) for some interval \(I\subset \R\) and \(g>0\).
    Let \(P\) be the spectral projection corresponding to \(\sigma_*\).
    Then
    \begin{equation*}
        A = -\I\,\commutator[\big]{H, \qliL_{H,g,\delta}(A)}
    \end{equation*}
    for all \(A\in\alg_{\Lat}\) satisfying \(A = P A P^\perp + P^\perp A P\), where \(P^\perp = \unit - P\).
    I.e.\ \(\qliL_{H,g,\delta}\) is the inverse of the Liouvillian \(A\mapsto-\I\,\commutator[\big]{H,A}\) for off-diagonal (w.r.t.\ \(P\)) \(A\).
    Moreover, if \(\diam{\sigma_*} < \delta\) then
    \begin{equation*}
        P \, \qliL_{H,g,\delta}(A) \, P
        =
        0
        \quadtext{for all}
        A\in\alg_{\Lat}
        .
    \end{equation*}
\end{proposition}

As a simple consequence (see~\cite[Corollary~4.2]{BRF2018}) \(\qlilgenerator(s) = \qliL_{H(s),g,\delta}\paren{\dot H(s)}\) indeed generates the spectral flow as a simple calculation shows
\begin{equation*}
    -\I\,\dot P
    = - \commutator[\big]{H, \qliL_{H,g,\delta}\paren[\big]{\dot P}}
    = - \qliL_{H,g,\delta}\paren[\big]{\commutator[\big]{H, \dot P}}
    = \qliL_{H,g,\delta}\paren[\big]{\commutator[\big]{\dot H, P}}
    = \commutator[\big]{\qliL_{H,g,\delta}\paren[\big]{\dot H}, P}
    ,
\end{equation*}
where we only used \cref{prop:qliL-general-properties} and off-diagonality of \(\dot P\) in the first step and the integral form of \(\qliL_{H,g,\delta}\) and \(\commutator{H,P}=0\) in all the other steps.

It is left to show that \(\qliL_{H(t),g,\delta}\paren{\dot H(t)}\) can be given by a polynomially decaying interaction.
Therefore, we approximate the individual terms using a conditional expectation.
For fermionic lattice systems, such a conditional expectation was constructed in~\cite{NSY2017}.
We collect its properties in the following~\namecref{lem:conditional-expectation}.

\begin{lemma}[{\cite[{Lemma~4.1, 4.2 and~4.3}]{NSY2017}}]
    \label{lem:conditional-expectation}
    Let \(X\subset\Lat\).
    Then there exists a unit-preserving, completely positive linear map \(\cexp{X}{}\colon \alg_{\Lat}\rightarrow\alg_{\Lat}\) satisfying
    \begin{thmlist}
        \item
            \(\cexp{X}{\algEven_{\Lat}}\subset\algEven_X\);
        \item \label{lem:cmean-E(ABC)=AE(B)C}
            \(\cexp{X}{ABC} = A \, \cexp{X}{B} \, C\) for all \(B\in\algEven_{\Lat}\) and \(A,C\in\algEven_{X}\);
            This in particular implies \(\cexp{X}{A}=A\) for all \(A\in\algEven_{X}\);
        \item
            \(\norm{\cexp{X}{}}=1\);
        \item
            \(\cexp{X}{} \circ \cexp{Y}{} = \cexp{X\cap Y}{}\), for \(X,Y\subset\Lat\);
        \item \label{lem:cmean-bound-for-difference-A-E(A)}
            If \(A\in\algEven_{\Lat}\) satisfies
            \begin{equation}
                \norm[\big]{\commutator{A,B}}
                \leq
                \eta \, \norm{A} \, \norm{B}
                \quadtext{for all}
                B\in\alg_{\Lat\setminus X}
                ,
            \end{equation}
            for some \(\eta>0\), then
            \begin{equation}
                \norm{A-\cexp{X}{A}} \leq \eta \, \norm{A}
                .
            \end{equation}
    \end{thmlist}
\end{lemma}

This allows to construct a decaying interaction representing \(\qliL_{H,g,\delta}(K)\).

\begin{proof}[Proof of \cref{lem:quasi-local-inverse-Liouvillian}]
    For the proof we first fix \(D\), \(\Aconst\), \(n\), \(\beta\), \(g\), \(\delta\), \(v_*\), and choose \(\alpha\) large enough (the possible values will be clear later on).
    In the following we allow the constant \(C\) to change in each step depending on all above-mentioned constants.
    Importantly, they do not depend on \(\Lambda\) nor \(\Phi\) directly.

    To prove the claim, we use the local decomposition technique:
    Each term \(\uts[s]{\Phi_\sltoperator(Z)}\) is split into a sum of terms in~\(\algEven_{Z_j}\) with norm decreasing rapidly in~\(j\), where~\(Z_j\) denotes the fattening defined in~\eqref{eq:definition-fattening}.

    For any \(O\in\alg_\Omega\) define
    \begin{align*}
        \Delta_0(O)
        &=
        \I \int_{\R} \Wfgd(s) \, \cexp[\big]{\Omega}{\uts[s]{O}} \diff s
    \shortintertext{and, for \(j\geq1\)}
        \Delta_j(O)
        &=
        \I \int_{\R} \Wfgd(s) \, \cexp[\big]{\Omega_j}{\uts[s]{O}} - \cexp[\big]{\Omega_{j-1}}{\uts[s]{O}} \diff s
        \\&=
        \I \int_{\R} \Wfgd(s) \, \cexp{\Omega_j}{}\circ
        \mathop{\paren[\big]{\unit - \cexp{\Omega_{j-1}}{}}}
        \paren[\big]{{\uts[s]{O}}}
        \diff s
        .
    \end{align*}
    Then \(\qliL_{H,g,\delta}(O)=\sum_{j=0}^\infty \Delta_j(O)\) where the sum is finite because \(\Lat\) is finite.

    For \(j=0\) we have the trivial bound
    \begin{equation*}
        \norm{\Delta_0(O)}
        \leq \norm{O} \, \norm{\Wfgd}_{L^1}
        .
    \end{equation*}
    For \(j\geq1\) and some \((\Ldim+1)/(\alpha+1)<\sigma<1\) to be chosen, \cref{thm:LR-bound-after-iteration} and the properties of the conditional expectation, yield
    \begin{align*}
        \Alignindent
        \norm*{
            \cexp{\Omega_j}{} \circ \mathop{\paren[\big]{\unit - \cexp{\Omega_{j-1}}{}}}
            \paren[\big]{{\uts[s]{O}}}
        }
        \leq
        \norm*{
            \paren[\big]{\unit - \cexp*{\Omega_{j-1}}{}}
            \paren[\big]{{\uts[s]{O}}}
        }
        \\&\leq
        2 \, \norm{O} \, \abs{\Omega} \, \paren[\Big]{
            \e^{\velocity \abs{s} - j^{1-\sigma}}
            + \sigmaconst \, (j+1)^{-\sigma\alpha} \, \velocity \abs{s} \, \paren[\big]{1+(\velocity \abs{s})^{\Ldim/(1-\sigma)}}
        }
    \end{align*}
    because \(\Ldist{\Omega,\Lat\setminus \Omega_{j-1}}=j\).

    Thus, for \(T=j^{p(1-\sigma)}/\velocity_*\) with \(p\in\intervaloo{0,1}\) we can bound
    \begin{align*}
        \Alignindent \norm*{
            \I \int_{-T}^T \Wfgd(s) \, \cexp[\Big]{\Omega_j}{}\circ
            \mathop{\paren[\big]{\unit - \cexp{\Omega_{j-1}}{}}}
            \paren[\big]{{\uts[s]{O}}}
            \diff s }
        \\&\leq
        4 \, \norm{O} \, \abs{\Omega} \, \norm{\Wfgd}_{\infty} \int_{0}^T
        \e^{\velocity s - j^{1-\sigma}}
        + \sigmaconst \, (j+1)^{-\sigma\alpha} \, \velocity s \, \paren[\big]{1+(\velocity s)^{\Ldim/(1-\sigma)}}
        \diff s
        \\&\leq
        4 \, \norm{O} \, \abs{\Omega} \, \norm{\Wfgd}_{\infty}
        \, \paren[\Big]{
            \e^{j^{p(1-\sigma)} - j^{1-\sigma}}
            + \sigmaconst \, (j+1)^{-\sigma\alpha} \, j^{2p(1-\sigma)} \, \paren[\big]{1+j^{p\Ldim}}
        }
        \\&\leq
        8 \, \norm{O} \, \abs{\Omega} \, \norm{\Wfgd}_{\infty}
        \, \paren[\Big]{
            \e^{j^{p(1-\sigma)} - j^{1-\sigma}}
            + \sigmaconst \, (j+1)^{-\sigma\alpha+2p(1-\sigma)+p\Ldim}
        }
        .
    \end{align*}
    Furthermore, for any \(m>0\) there exists \(C_m\) such that
    \begin{align*}
        \Alignindent \norm*{
            \I \int_{\abs{s}>T} \Wfgd(s) \, \cexp[\Big]{\Omega_j}{}\circ
            \mathop{\paren[\big]{\unit - \cexp{\Omega_{j-1}}{}}}
            \paren[\big]{{\uts[s]{O}}}
            \diff s
        }
        \\&\leq
        2 \, \norm{O} \int_{\abs{s}\geq T} \Wfgd(s) \diff s
        \\&\leq
        8 \, \norm{\Wf_{g,\delta}}_{\infty}
        \, C_m \, \norm{O} \, \velocity_*^m \, j^{-mp(1-\sigma)}
    \end{align*}
    by the properties of \(\Wfgd\).
    Hence, altogether we find
    \begin{equation*}
        \norm{\Delta_j(O)}
        \leq
        8 \, \norm{\Wf_{g,\delta}}_{\infty} \, \norm{O} \, \abs{\Omega}
        \, \paren[\Big]{
            \e^{j^{p(1-\sigma)} - j^{1-\sigma}}
            + \sigmaconst \, (j+1)^{-\sigma\alpha+2p(1-\sigma)+p\Ldim}
            + C_m \, \velocity_*^m \, j^{-mp(1-\sigma)}
        }
        .
    \end{equation*}

    An interaction for \(A:=\qliL_{H,g,\delta}(\sltoperator)\) is given by
    \begin{equation*}
        \Phi_A(Z)
        =
        \sum_{j=0}^\infty \sumstack{Y\subset\Lat\suchthat\\Y_j=Z} \Delta_j(\Phi_\sltoperator(Y))
        .
    \end{equation*}
    It follows that
    \begin{align*}
        \sumstack{Z\subset\Lat\suchthat\\z\in Z} \frac{\abs{Z}^n \, \norm{\Phi_A(Z)}}{F_\beta\pLdiam{Z}}
        &\leq
        \sumstack{Z\subset\Lat\suchthat\\z\in Z} \sum_{j=0}^\infty \sumstack{Y\subset\Lat\suchthat\\Y_j=Z} \frac{\abs{Z}^n \, \norm{\Delta_j(\Phi_\sltoperator(Y))}}{F_\beta\pLdiam{Z}}
        \numberthis{eq:proof-qliL-interaction-norm-of-generator}
        \\&\leq
        \sum_{j=0}^\infty \sum_{Y\subset\Lat} \unit_{z\in Y_j} \frac{\abs{Y_j}^n \, \norm{\Delta_j(\Phi_\sltoperator(Y))}}{F_\beta\pLdiam{Y_j}}
        .
    \end{align*}
    The \(j=0\) term is bounded by \(\norm{\Wfgd}_{L^1} \, \norm{\Phi_\sltoperator}_{\beta,n}\).
    For \(j\geq1\) and \(z\in Y_j\), there exists \(y\in B_z(j)\cap Y\).
    Moreover, \(\abs{Y_j} \leq \abs{Y} \, \Vconst \, (j+1)^\Ldim\) and \(\Ldiam{Y_j} \leq 2j + \Ldiam{Y}\).
    Hence, the rest of the sum is bounded by
    \begin{align*}
        \Alignindent
        \sum_{j=1}^\infty \sum_{y\in B_z(j)} \sumstack{Y\subset\Lat\suchthat\\y\in Y}
        \begin{multlined}[t]
            \frac{\abs{Y}^n \, \Vconst^n \, (j+1)^{n\Ldim} \, \norm{\Phi_\sltoperator(Y)} \, \abs{Y}}{F_\beta\pLdiam{Y} \, F_\beta(2j)}
            \\\times 8 \, \norm{\Wf_{g,\delta}}_{\infty}
            \, \paren[\Big]{
                \e^{j^{p(1-\sigma)} - j^{1-\sigma}}
                + \sigmaconst \, (j+1)^{-\sigma\alpha+2p(1-\sigma)+p\Ldim}
                + C_m \, \velocity_*^m \, j^{-mp(1-\sigma)}
            }
        \end{multlined}
        \\&\leq
        \begin{aligned}[t]
            &2^{3+\beta} \, \, \norm{\Wf_{g,\delta}}_{\infty} \, \Vconst^n \, \norm{\Phi_\sltoperator}_{\beta,n+1}
            \\&\times
            \sum_{j=1}^\infty
            \sumstack[r]{y\in B_z(j)}
            \, (j+1)^{\beta+n\Ldim}\,
            \paren[\Big]{
                \e^{j^{p(1-\sigma)} - j^{1-\sigma}}
                + \sigmaconst \, (j+1)^{-\sigma\alpha+2p(1-\sigma)+p\Ldim}
                + C_m \, \velocity_*^m \, j^{-mp(1-\sigma)}
            }.
        \end{aligned}
    \end{align*}
    The remaining sum over \(y\) can be bounded by \(\abs{B_z(j)} \leq \Vconst\, (j+1)^\Ldim\).
    The infinite sum over the first term in the parenthesis, is then bounded for \(p,\sigma\in\intervaloo{0,1}\).
    The same holds for the last term by choosing \(m\) appropriately.
    The sum over the second term is bounded whenever \({\sigma\,\alpha > 1 + (n+1)\,\Ldim + \beta + 2p\,(1-\sigma)+p\Ldim}\), i.e.\ as long as \({\alpha>(n+1)\,\Ldim+1+\beta}\) by choosing \(\sigma\) appropriately.
    The final constant thus depends on \(\Ldim\), \(\alpha\), \(\beta\), \(n\) (after choosing \(\sigma\) and \(p\) optimally), but also on the \(\velocity_*\) and \(g\) and \(\delta\) and thus directly on the norm of the interaction \(\Phi\) and its spectral properties.
\end{proof}

Proving \cref{cor:LPPL} uses very similar ideas, but is simpler because we only consider strictly localized \(\dot H(s)\in \alg_X\) and far apart observables.

\begin{proof}[Proof of \cref{cor:LPPL}]
    Let \(W(s) = \sum_{Z\subset X} \dot \Phi(Z,s)\), then the generator of \(U\) from \cref{thm:automorphic-equivalence} is \(\qlilgenerator=\qliL_{H,g,\delta}(W)\).
    For the proof, we approximate \(U(s)\) by a unitary \(V_Y(s)\), which acts like the identity on \(\alg_Y\), where \(B\) is supported.
    Therefore, let \(V_Y(s)\) be defined by its generator
    \begin{equation}
        \label{eq:proof-LPPL-localized-generator}
        \qlilgenerator_Y(t)
        :=
        \I \int_{\R} \Wfgd(s) \, \cexp[\Big]{\Lambda \setminus Y}{\uts[s]{}\paren[\big]{W(t)}} \diff s
        .
    \end{equation}
    As in the proof of \cref{lem:quasi-local-inverse-Liouvillian}, and abbreviating \(r=\Ldist{X,Y}\), we obtain
    \begin{equation*}
        \norm[\big]{\qlilgenerator_Y(t) - \qlilgenerator(t)}
        \leq
        C \, \norm{W(t)} \, \abs{Y}
        \, \paren[\Big]{
            \e^{(r+1)^{p(1-\sigma)} - (r+1)^{1-\sigma}}
            + (r+2)^{-\sigma\alpha+2p(1-\sigma)+p\Ldim}
            + (r+1)^{-mp(1-\sigma)}
        }
        ,
    \end{equation*}
    using the Lieb-Robinson bound from \cref{thm:LR-bound-after-iteration}, which requires a uniform bound on \(\norm{\Phi}_{\alpha,1}<\infty\) for some \(\alpha>D\).
    We can further bound this by \(C \, \norm{W(t)} \, \abs{Y} \, F_{\alpha-\varepsilon}(r)\) for all \(\varepsilon>0\) by choosing \(p,\sigma\in\intervaloo{0,1}\) and \(m\) appropriately.
    By the fundamental theorem of calculus,
    \begin{equation*}
        U(s) - V_Y(s)
        =
        - U(s) \int_0^s \odv*{U(t)^*\,V_Y(t)}{t} \diff t
        =
        \I \, U(s) \int_0^s U(t)^*\,\paren[\big]{\qlilgenerator(t)-\qlilgenerator_Y(t)}\,V_Y(t) \diff t
        .
    \end{equation*}
    And hence \(
        \norm[\big]{U(s)-V_Y(s)}
        \leq
        s \, \sup_{t\in I} \, \norm[\big]{\qlilgenerator(t) - \qlilgenerator_Y(t)}
    \) for all \(s\in I\).
    Since \(V_Y(s)\in \algEven_{\Lambda\setminus Y}\), it satisfies \(\commutator{V_Y(s),A} = 0\).
    Hence, after using cyclicity of the trace, we have
    \begin{align*}
        \abs[\big]{ \trace[\big]{P(s)\,A} -\trace[\big]{P(0)\,A} }
        &=
        \abs[\big]{ \trace[\big]{P(0)\,U(s)^*\,\commutator[\big]{A,U(s)-V_Y(s)}}}
        \\&\leq
        2 \, \trace[\big]{P(0)} \, \norm{A} \, s \, C \, \sup_{t\in I} \, \norm{W(t)} \, \abs{Y} \, F_{\alpha-\varepsilon}\pLdist{X,Y}
        ,
    \end{align*}
    and conclude~\eqref{eq:cor-LPPL-trace-bound-norm}.

    To obtain~\eqref{eq:cor-LPPL-trace-bound-interaction-norm}, we split \(W(t) = \sum_{Z\subset X} \dot \Phi(Z,t)\) in~\eqref{eq:proof-LPPL-localized-generator} and apply the localization to each summand.
    Following the remaining arguments, we then obtain
    \begin{align*}
        \Alignindent
        \abs[\big]{ \trace[\big]{P(s)\,A} -\trace[\big]{P(0)\,A} }
        \\&\leq
        2 \, \trace{P(0)} \, \norm{A} \, s \, C \, \sup_{t\in I} \, \sum_{Z\subset X} \, \norm{\dot \Phi(Z,t)} \, \abs{Z} \, F_{\alpha-\varepsilon}\pLdist{Z,Y}
        \\&\leq
        2 \, \trace{P(0)} \, \norm{A} \, s \, C
        \, \sum_{z\in X}
        \sum_{y\in Y}
        \, \sup_{t\in I}
        \, \sumstack[r]{Z\subset X\suchthat\\z\in Z}
         \, \norm{\dot \Phi(Z,t)} \, \abs{Z} \, F_{\alpha-\varepsilon}\pLdist{z,y}
        \\&\leq
        2 \, \trace{P(0)} \, \norm{A} \, s \, C
        \, \Aconst \, \tfrac{\alpha-\epsi-D+1}{\alpha-\epsi-D} \, \abs{Y}
        \, \norm{\dot \Phi}_{0,1} \, F_{\alpha-D-\varepsilon}\pLdist{X,Y}
        ,
    \end{align*}
    if \(\alpha-\epsi>D\), where similarly to~\eqref{eq:example-bound-of-LRsum} we used that for \(\beta>D\) and \(R \geq 1\)
    \begin{equation*}
        \sumstack[r]{z\in \Lambda\suchthat\\ \Ld{z,y}\geq R} F_{\beta}\pLd{z,y}
        \leq
        \Aconst \, \paren[\bigg]{
            F_{\beta}(R) + \int_{R}^\infty F_{\beta-D+1}(r) \diff r
        }
        \leq
        \Aconst \, \tfrac{\beta-D+1}{\beta-D} \, F_{\beta-D}(R)
        .
        \qedhere
    \end{equation*}
\end{proof}

\printbibliography[heading=bibintoc]{}

\end{document}